\documentclass[12pt, onecolumn
, floatfix, amsmath,amssymb,amsfonts, aps,pre, 
]{revtex4-2}
\usepackage{graphicx}
\usepackage{dcolumn}
\usepackage{bm}
\usepackage{multirow}
\usepackage{color}
\usepackage{ulem}
\usepackage{float}
\usepackage{cancel}
 \usepackage{amssymb}

\begin{document}


\title{Magnetized granular particles running and tumbling on $S^{1}$
}

\author{M. Ledesma‑Motolinía, J. L. Carrillo‑Estrada}
 \affiliation{Instituto de Física “Luis Rivera Terrazas”, Benemérita Universidad Autónoma de Puebla, Puebla 72570,Mexico.}
\author{A. Escobar, F Donado,}%
 
\affiliation{%
 Instituto de Ciencias B\'asicas e Ingenier\'ia de la Universidad Aut\'onoma del Estado de Hidalgo-AAMF, Pachuca, 42184, Hgo., M\'exico.}
\author{Pavel \surname{Castro-Villarreal}}
\email{corresponding author:pcastrov@unach.mx }
\affiliation{Facultad de Ciencias en F\'isica y Matem\'aticas,
Universidad Aut\'onoma de Chiapas, Carretera Emiliano Zapata, Km. 8, Rancho San Francisco, 29050 Tuxtla Guti\'errez, Chiapas, M\'exico}

\begin{abstract}

It has been shown that a nonvibrated magnetic granular system, when it is feeded by means an altenating magnetic field, behaves with most of the distinctive physical features of active matter systems. 
In this work we focus our attention on
the simplest granular system composed 
by a single magnetized spherical particle allocated in a quasi one-dimensional circular channel that 
receives energy from a  magnetic field reservoir 
and transduces it into a running and tumbling motion
. The theoretical analysis 
based 
on the run and tumble model on a circle of radius $R$ forecasts the existence of a dynamical phase transition between an erratic motion (disordered phase) when the characteristic persistence length of the run and tumble motion, $\ell_{c}<R/2$, to a 
persistent motion (ordered phase) when $\ell_{c}>R/2$
. 
It is found  that the limiting behaviours of these 
phases correspond to a Brownian motion on the circle and a simple uniform circular motion, respectively. 
It is qualitatively shown that the lower magnetization of a particle, the larger persistence lenght is. It is so at least within the experimental limit of validity of our experiments. Our 
results show 
a very good agreement between theory and experiment. 

\end{abstract}

\maketitle

\section{Introduction}

 Active matter is a term conceived to classify those systems composed of entities (called {\it active particles}) that can extract free-energy from a reservoir to transform it into kinetic energy; it makes these particles transform into self-propelled units. The available reservoir can be on-board or environmental. The energy consumed by each active particle fuels an intrinsic mechanism that, in dissipating, transduces in a type of systematic motion that is generally common for all active particles \cite{MarchettiRMP2013, Bechinger2016, Fodor2018}. The active matter can have a variety of patterns depending upon its constituents that generate emergent and collective non-equilibrium phenomena \cite{Ramaswamy2017, Bowick2022} such as the marvellous dancing of birds flock in the sky \cite{CavagnaPRSB2013} (other examples can be seen at \cite{VicsekPhysRep2012}).  This type of matter can be presented also at the sub-cellular scale \cite{TailleurPRL2008, TailleurEPL2009, Wang2011}, and it can have non-biological components \cite{Ramaswamy2017}.

The constituents of active matter might be of a very diverse origin, from biological entities such as bacteria, unicellular protozoa, spermatozoa, among many others, to man-made experimental realizations like artificial microswimmers that mimicrises the previous biological microorganism such as Janus particles, colloidal propellers, Pt-loaded stomatocytes, water droplets, among many others \cite{Ebbens2010-SoftMatter, Bechinger2016}. One the one hand, some of these examples show that self-propulsion can be carried out by conversion between chemical to mechanical energy. For instance, by coating with platinum a hemisphere of a polystyrene sphere immersed in a bank of hydrogen peroxide allows to achieve a transition between directed motion to a Brownian motion  by means of the platinum catalyst processes (this is a typical example of self-diffusiophoresis) \cite{HowsePRL2007}. Another illustrative example, is a particle half-metal coated under laser irradiation that undergoes  self-thermophoresis  due to a local temperature gradient induced by a laser beam \cite{JiangPRL2010}. On the other hand, granular materials turn out to be examples of physical realization of active matter \cite{Yamada2003, Narayan2007, Cheng2021}, where  the shape anisotropy of the single grains under vertical vibration can be produce motility on a horizontal surface \cite{Yamada2003}. For instance, the single-particle active motion can be generated in a vibrated granular system, where the main features of the Active Brownian Motion model has been proved in \cite{LeeWalsh2017}.   Furthermore, recent experimental studies on the dynamics of active granular particles have been performed using ‘‘micro robotic creatures’’, a robot toy called ``Hexbugs Nano" with an internal motor that produces vibrations, where the most outstanding properties of active motion are described through a Langevin stochastic model whence inertial effects are highlighted \cite{Tapia_Ignacio_2021}.



In contrast to the above vibrated granular systems, a nonvibrating granular system composed of many metallic balls under an alternating magnetic field is a genuine $2D$ active matter system. Unlike a vibrated one, in which particles can move vertically, in a nonvibrating granular system particles always remain in contact with the lower surface of the cell. This 2D granular system exhibits the typical features of other active matter systems, including the dependence of the collective behaviour on the confinement conditions \cite{Deseigne2012} and the particle concentration, as well as the merging correlations that originate self-organization phenomena like flocking, arrested structures, and crystallization \cite{Cecilio2016, Cecilio2020, Monica2021}. The motion of particles during crystallization phenomena in confined conditions experiments change from diffusive to subdiffusive. In particular, it has been observed that confinement fastens the crystallization process \cite{Meldrum2020, Jung2020},   as well it induces complex correlations that, under certain conditions, generate emergent phenomena like vortices, and flocking \cite{Bechinger2016}. 

Here,  we wish to understand some of the main aspects of the stochastic dynamics of active granular matter confined in a quasi-1D-channel such as a circular track. We study this granular system based on a magnetized spherical bead activated by an alternating magnetic field used as a reservoir. The particle takes energy from the magnetic field and transforms it into kinetic energy through the following mechanism. In the presence of the magnetic field, the magnetic dipole of the particle tries to align with the field to minimize the energy of the system. When the magnetic field reverses its direction, the particle rotates to align again with the field. In this process, a particle rotates rolling over the surface because of gravity and friction. In this manner, the particle moves along the circular channel. When the field reverses direction again, a new impulse acts over the particle making it continue rolling in the same direction or reversing the direction inside the channel to a random new direction leading to an active motion. If confinement were not present, the particle would change its motion to a random new direction leading to a 2D-active behaviour \cite{Donado2017}.

One of the main aims of the present work is to understand how well the particle behaviour of this nonvibrating $1D$-granular-system can be described in terms of the run-and-tumble model  \cite{MartensEPJE2012, Soto2014,Sepulvedad2016,Barrius02021} on the circle $S^{1}$. In the Run and Tumble model, the constraint to the line turns out in the continuous persistent random walk model introduced by S. Goldstein \cite{GoldsteinQJMAM1951} back to the fifties, where the limiting process of the probability to find a particle, among a large number of non-interacting particles, that moves to the left or the right with equal probabilities and with a uniform velocity is described exactly by a solution of a telegrapher equation (TE). This equation has also arisen as just an approximation within the standard active Brownian motion model \cite{Romanczuk2012} studied in $2D$ flat space \cite{Sevilla2014, SevillaPRE2015} and in $2D$ curved surface \cite{CastroPRE2018}; and also in the Run and Tumble model for active particle \cite{tailleur2008, CatesEPL2013}. However, for the $1D$ situation, the telegrapher's equation description is exact, as opposed to the higher dimensions \cite{SevillaCastro2021}, thus its solutions have the full description of the active motion behaviour.  In particular, in contrast to a higher dimension, where the hydrodynamic-like description of the active dynamics consist of an infinite tower of a hierarchy of equations \cite{SevillaCastro2021, Kurzthaler2017, KurzthalerPRL2018}, in the  $1D$ case consist of on the continuity equation and a current probability equation, namely, 
\begin{eqnarray}
\frac{\partial}{\partial t}\rho(s,t)&=&-\frac{\partial}{\partial s}\mathbb{J}\left(s,t\right),\label{ContEq}\\
\frac{\partial }{\partial t}\mathbb{J}\left(s,t\right)&=&-\frac{1}{\tau_{c}}\mathbb{J}\left(s,t\right)-v_{0}^{2}\frac{\partial}{\partial s}\rho\left(s,t\right),
\label{CurrentEq}
\end{eqnarray}
where $\tau_{c}$ is the average time elapsed before the particle performs a
tumble, $v_{0}$ is the constant average persistence velocity and $s$ is the arc-length of the circle. The quantity $\rho(s,t)$ represents the probability density function to find a particle at position $s$ after a time $t$ has passed, and $\mathbb{J}\left(s,t\right)$ is the current probability function that describe preferential probability direction of the particle motion. The theoretical predictions of the Run and Tumble model on $S^{1}$, described by (\ref{ContEq}) and (\ref{CurrentEq}), predicts the existence of a critical persistence length $\ell_{c}=v_{0}\tau_{c}$ that distinguishes between  two-state of motion, that is, the model predicts a dynamical phase transition between an erratic motion (disorder phase) to a persistent motion (order phase), where the extreme behaviour of these states correspond to a Brownian motion on the circle and to a simple uniform circular motion, respectively. A good agreement between theory and experiment within the parameter regime considered in the experiment shall be shown below. 

Our paper is organized as follows: In Sec. \ref{SectII} the experimental setup to study the magnetic granular active motion in $1D$ is introduced; in particular, there it is described the main features of the stochastic motion of a magnetized ball confined on a circular channel and subjected to the time-varying magnetic field used as a thermal reservoir. In Sec. (\ref{SectIII}) is developed the Run and Tumble model defined on a circle $S^{1}$; in particular, there is computed exactly the probability density function $\rho(s,t)$, the current probability function $\mathbb{J}(s,t)$ as well as expectation values of different physical quantities such as the mean-squared displacement.  In Sec. \ref{SectIV} we give a comparison between the experimental results and the theoretical predictions made by the RTM model on $S^{1}$. Here, we describe the most salient characteristic of a phase transition that undergo the particle's motion between an erratic motion and a soft motion. Finally, in Sec. \ref{SectV} we give our concluding remarks and perspectives.

\section{Experimental setup}\label{SectII}

\begin{figure}[ht]
\begin{center}
    \includegraphics[scale=0.6]{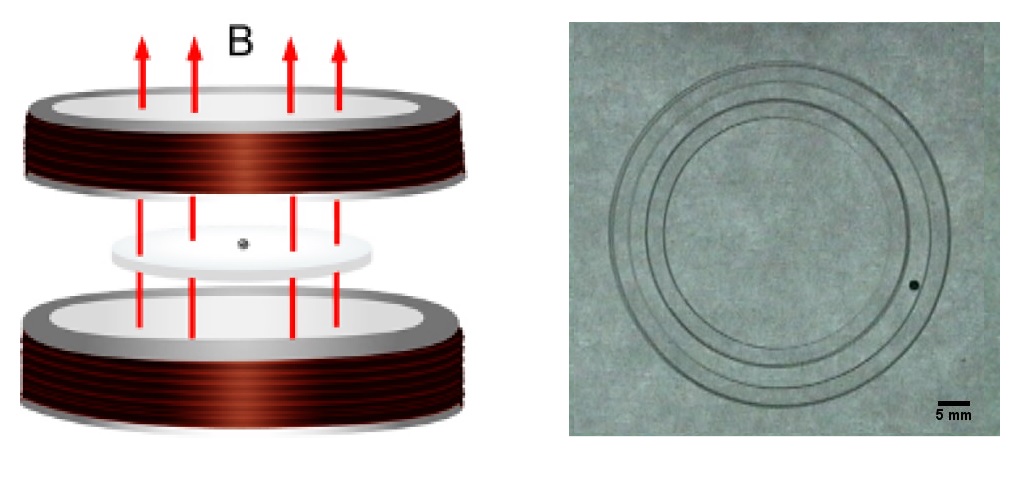}~~~~~~ 
    \caption{{Left: Experimental arrangement consists of a pair of Helmholtz coils that produce an oscillating magnetic field $B$ through the observation cell. Right: A lithographic circular channel with a width value of $1.6~{\rm mm}$ and one metallic bead of diameter $\sigma=1~{\rm mm}$. The channel radius is measured from the centre of the channel to the centre of the ball. The scale bar is $5~{\rm mm}$.}}
    \label{1}
\end{center}
\end{figure}

In this work, the experimental arrangement described in recent works \cite{Cecilio2020, Escobar2021, Torres2020, Miranda2019} is slightly modified to study the stochastic motion of a metallic ball confined to a lithographic circular channel made on an acrylic plate. Briefly speaking, the setup consists of a pair of Helmholtz coils, where a flat and horizontal plate, called observation cell, is placed in the homogeneous magnetic field region as shown schematically in the left of the figure (\ref{1}). A closed circular channel  is built on the observation cell, in such a way that the particle motion becomes confined into a quasi-1D-channel (see right of the figure (\ref{1})). In the present experiment, it is  analysed the motion of a single steel bead with 1 mm of diameter. The coils are powered by a KEPCO brand power source controlled by a National Instruments data acquisition card through a sinusoidal signal generated by a homemade program in LabView. The coil system generates an oscillatory magnetic field of the form $B=B_o \sin(2 \pi f t)$. The frequency $f$ is kept constant at 9.24 Hz. Recall that previously it has been shown that in this kind of systems the amplitude of the magnetic field $B_{o}$ can be interpreted as an effective temperature, up to a proportional coefficient \cite{Cecilio2016}. Therefore, for simplicity, here the effective temperature is denoted by $B_o$, quantified in Gauss units denoted by ${\rm G}$. 

 The experiments reported here start by putting a single particle in the quasi-1D-channel and turning on the alternating magnetic field. In order to search for the influence of the particle magnetization on the stochastic motion, it is prepared particles with magnetic dipole $m_{1}$, obtained after exposing the particle under the magnetic field of magnitude $66 {\rm G}$ for one hour, and particles with magnetic dipole $m_{2}$, obtained by using a magnetic field of magnitude $100 {\rm G}$ for the same time.  Afterwards,  it is calculated  the Mean-Square Euclidean Displacement (MSD) in order to determine the stochastic motion behaviour for particles with $m_{1}$ and $m_{2}$, respectively.  The MSD is  defined by $\left<\Delta {\bf r}^{2}(t)\right>=\left<|{\bf r}(t)-{\bf r}(0)|^2\right>$, where the displacement $\Delta {\bf r}$ is measured in particle's diameter units $\sigma$, and ${\bf r}(t)$ is the vector position measured from the center of the circular channel to the center of the metallic ball. 
 
We carried out four series of experiments. In the series $S_1$, we used a particle with magnetic moment $m_1$, a magnetic field amplitude $B_{o}=55\,{\rm G}$ and a set of various radius of the circular channel $R$, which is measured in units of the particle's diameter, $\sigma$. The ratio $R/\sigma$ ranges from $5$ to $34$. For the series $S_2$, we used a particle with magnetic moment $m_2$ at the same values of  $B_{o}$, and $R$ as in $S_1$. In the series $S_3$, we used $m_1$, circular track radius $R=20~\sigma$, and varied the effective temperature, $B_{o}$ from $44$ to $88\,{\rm G}$. For the series $S_4$, we used $m_2$, circular track radius $R=21~\sigma$, and varied $B_{o}$ from $11$ to $88~ {\rm G}$. In the series $S_3$ we observed at the lower effective temperatures, $11~{\rm G}$, $22~{\rm G}$, and $33~{\rm G}$, that the particle tend to adhere to the wall and stop moving. Also, at higher effective temperatures in $S_3$ and $S_4$, the response of the particle is no longer proportional to the amplitude of the field, such that the amplitude can not be used any longer as an effective temperature. 

The experiments were recorded using a CCD video camera at a speed of $30$ frames per second. Each experiment lasted $5$ minutes, which is enough time before the balls modify their magnetization.  The analysis of the videos was done by using the ImageJ package. 

\begin{figure}[ht]
\begin{center}
   \includegraphics[scale=0.5]{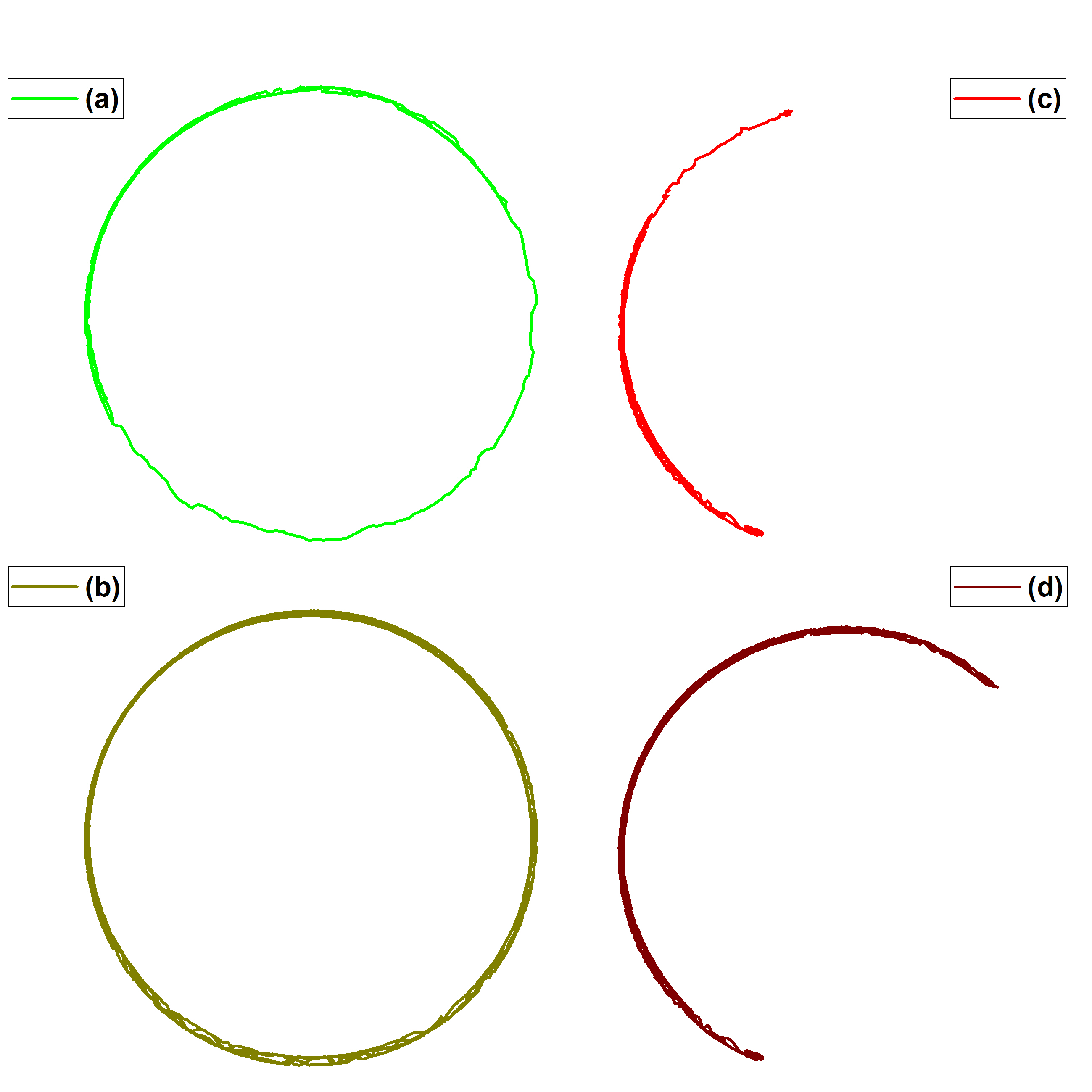}
    \caption{{(a) Particle trajectories at 20 seconds and (b) 66 seconds for series S1, (c) particle trajectories at 20 s and (d) 66 s for series S2. In both series, the radius of the circular track channel is $R=34~\sigma$. 
}}
 \label{3}
\end{center}
\end{figure}

Figure  (\ref{3}a) and figure  (\ref{3}b) show the particle trajectories in a circular channel with  $R=34~\sigma$ for series $S_1$ at 20 seconds and 66 seconds, respectively. At the same time intervals, figure  (\ref{3}c) and figure (\ref{3}d) show the particle trajectories for series $S_2$. At the shorter time, it is observed that the particle in series $S_1$ exhibits larger traces and a larger displacement than the particle in the conditions of series $S_2$, clearly, this particle travels a smaller portion of the entire circular channel.


\begin{figure}[ht]
\begin{center}
    \includegraphics[scale=0.5]{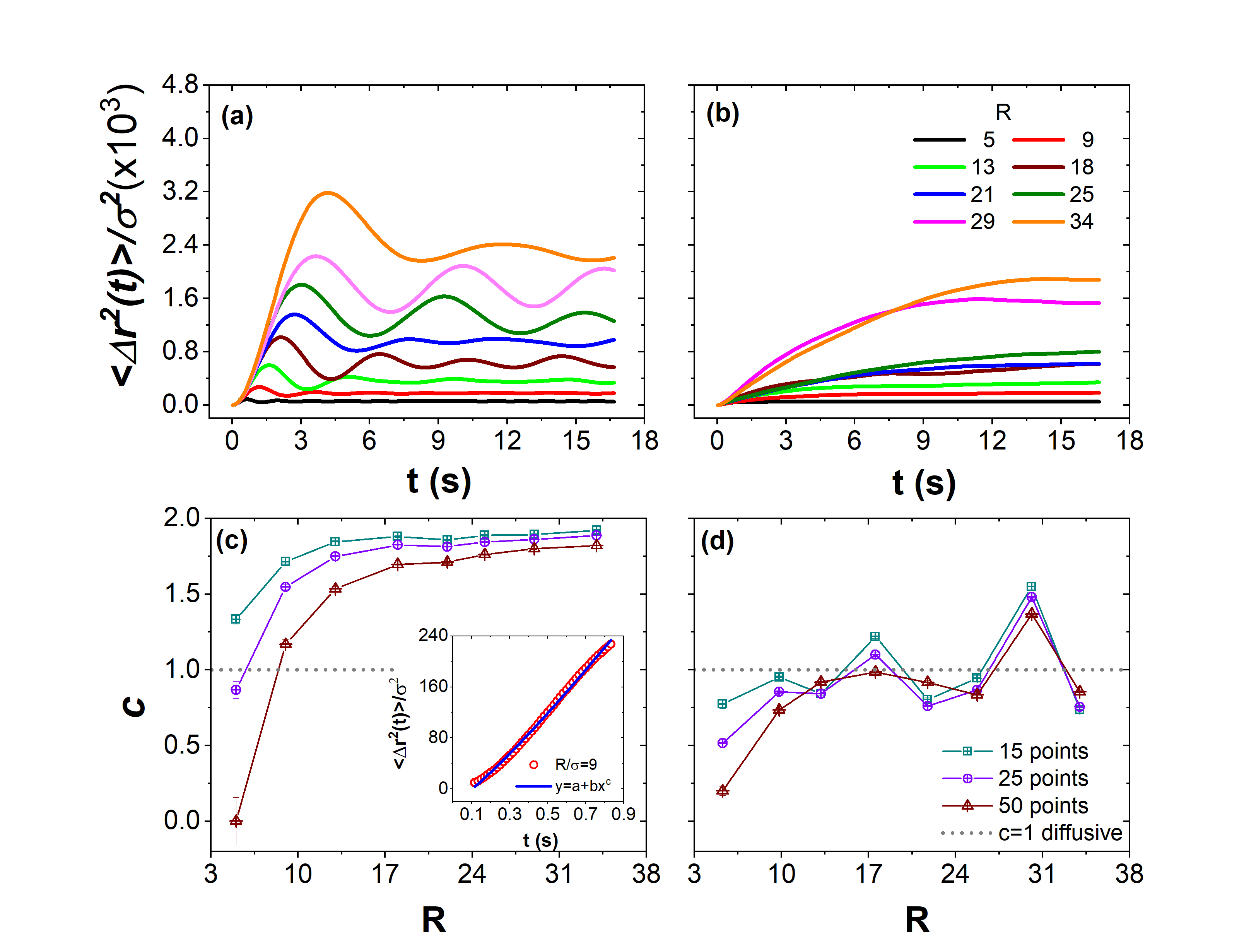}
    \caption{{Mean-square displacement for (a) series experiment S1, and (b) series experiment S2 for several radius $R$. Exponent $c$ of the power law $\left<\Delta {\bf r}^{2}(t)\right>\propto t^{c}$, at the short-time regime of the curves a) and b), as a function of the radius $R$; for (c) series S1, and (d) series S2. Inset: Linear adjustment of the first segment of red curve from (a) to calculate the exponent $c$. The exponent $c$ is calculated taking the first $15$, $25$ and $50$ from the MSD, respectively.}}
    \label{4}
\end{center}
\end{figure}

Figure (\ref{4}a), shows the mean-squared displacement curves of the series experiment $S_1$. It is observed there that each curve reaches a maximum value, and then oscillates. The general trend is that the amplitude of the oscillations increases as long as $R$ increases. Also, the maximum values appear to shift to the right as $R$ increases. Both of these results can be explained considering the accessibility to greater displacements when $R$ increases. Figure (\ref{4}b) shows the corresponding curves from the series $S_2$. These curves, which were obtained using a particle with larger magnetization, do not exhibit oscillations. They  grow monotonically toward a limit value. A direct explanation of this phenomena can be obtained by observing the role of the magnetization of the particle as shown in Figure (\ref{3}). The initial portion of each set of MSD curves was fitted using power-law functions in order to figure out a qualitative behaviour of the stochastic motion.   Figure (\ref{4}c) shows the exponents $c$ obtained from each power-law function, as a function of the radius. It is clear that the behaviour of the exponent corresponding to the series $S_1$ exhibits the same general trend for all the $R$ values. The exponent indicates that the motion is superdiffusive with an exponent roughly of $\sim 1.8$, considering a radius above $21~\sigma$. Figure (\ref{4}d)  shows the corresponding $c$ values as function of radius. It is observed the average value is around the unity, indicating that for this magnetization of the particle, its motion is nearly diffusive.

\begin{figure}[ht]
\begin{center}
   \includegraphics[scale=0.5]{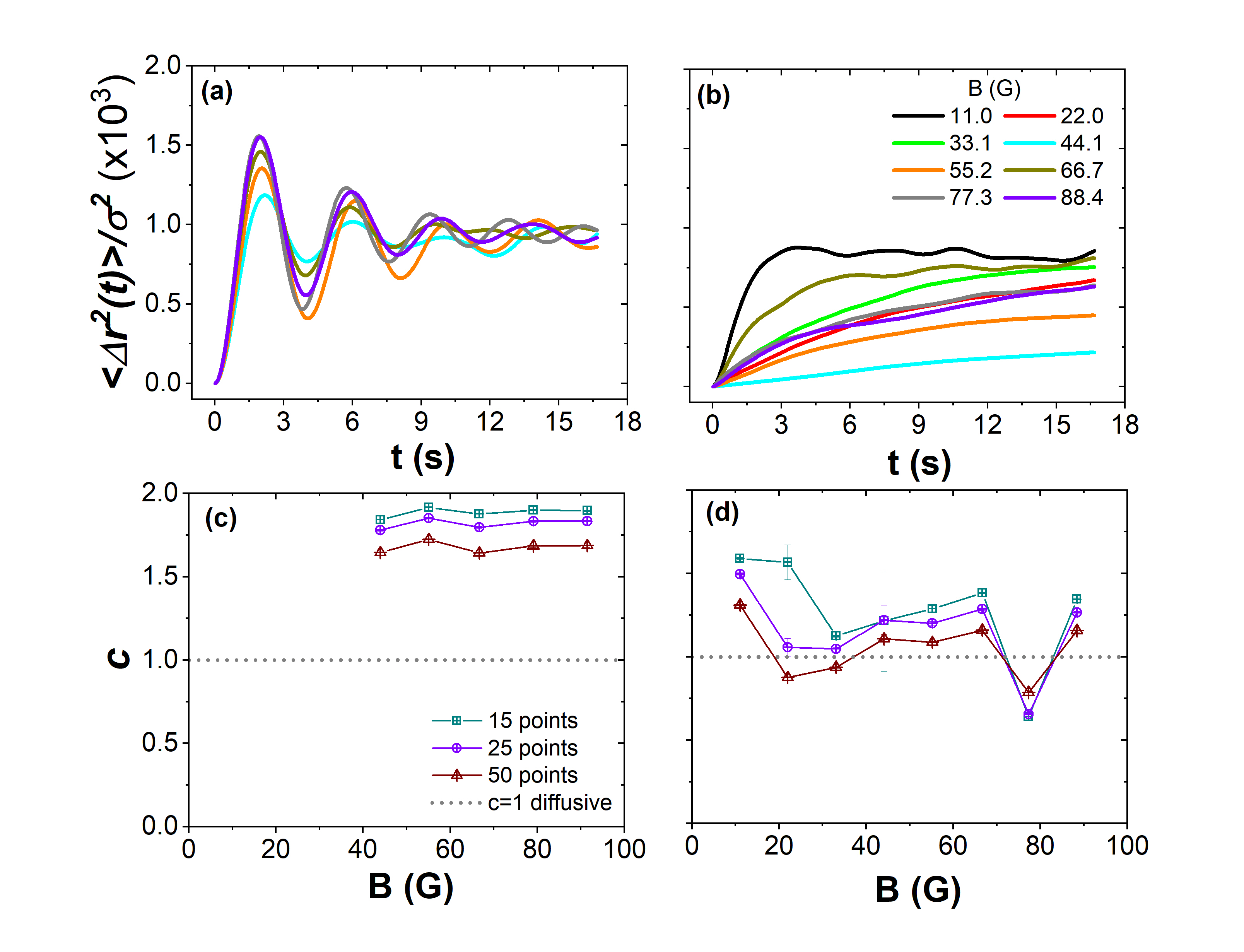}
    \end{center}
    \caption{Mean-square displacement for (a) series S3, and (b) series S4 as a function of the amplitude of magnetic field B. Exponent $c$ of the power law $\left<\Delta {\bf r}^{2}(t)\right>\propto t^{c}$, at the short-time regime of the curves a) and b), as a function of the radius $R$; magnetic field B for series $S_3$ (c), and series $S_4$  (d). The exponent $c$ is calculated taking the first $15$, $25$ and $50$ from the MSD, respectively. }
    \label{5}
\end{figure}

Figure (\ref{5}a), shows the mean-squared displacement curves  obtained from the series $S_3$, here we varied the effective temperature $B_{o}$. As in series $S_1$, each curve reaches a maximum value and then oscillates. The general trend is that the amplitude of the oscillations increases as $B_{o}$ increases. Also, the maximum values slightly shifts to the left as $B_{o}$ increases. Figure (\ref{5}b) shows the corresponding curves of series $S_4$. These curves, which were obtained using a particle with larger magnetization, grow almost monotonically toward a limit value, presenting superimposed small oscillations.  The exponents $c$ obtained from each power-law function is shown in figure (\ref{5}c), as a function of the effective temperature $B_{o}$. We observe here that although the exponent depends on the number of data used to fit, the general trend is the same. The motion is superdiffusive with an exponent around $\sim 1.8$. Figure (\ref{5}d) shows the corresponding $c$ values as a function of effective temperature. In these conditions the particle motion oscillates between superdiffusive to subdiffusive with an average value above the unit indicating that the motion is also superdiffusive, but less than in series $S_3$.

\section{Theory of the Run and Tumble motion on the circle}\label{SectIII}

In this section, we developed the run and tumble model for a particle confined to move on a compact planar Jordan curve $\gamma$ of perimeter $L$. In the simplest case of a one dimensional manifold $\gamma$, the possible states of an active particle are the motion to the right  or to the left. Mathematically, these states can be characterized as elements of a  $0$-dimensional sphere $S^{0}=\{-1,1\}$ accounting for the two directions of the particle  velocity $\hat{v}=\pm 1$.  Following the general model of {\it run and tumble} \cite{MartensEPJE2012} for a particle in a $d-$dimensional curved manifold (\ref{RTcurvedspace}) (see the appendix \ref{sec:sample:appendix}), it is straightforward to get the Goldstein's equations on $\gamma$ \cite{GoldsteinQJMAM1951}, 
\begin{eqnarray}
\frac{\partial}{\partial t}P_{+}(s, t)+v_{0}\frac{\partial }{\partial s} P_{+}(s, t)&=&-\lambda P_{+}(s,  t)+\frac{\lambda}{2} \sum_{k=\pm} P_{k}(s, t), \nonumber\\
\label{model1}\\
\frac{\partial}{\partial t}P_{-}(s, t)-v_{0}\frac{\partial }{\partial s} P_{-}(s, t)&=&-\lambda P_{-}(s,  t)+\frac{\lambda}{2} \sum_{k=\pm} P_{k}(s, t),\nonumber\\
\label{model2}
\end{eqnarray}
where $P_{\pm}(s, t)=P(s, \hat{v}=\pm1, t)$ represents the probability density function to move either to the right $(+1)$ or to the left $(-1)$,   $v_{0}$ is the constant average particle velocity, and $\lambda$ is a uniform tumbling frequency rate related to the persistent time defined by $\tau_{c}=\lambda^{-1}$, which gives the average time elapsed before the particle performs a tumble. Similarly,  we also define the persistent length by $\ell_{c}=v_{0}\tau_{c}$, meaning the average length run by the particle during the persistent time. One can use the radius of the circle $R$ in order to define a large persistent length as $\ell_{c}\gg R$ and small persistent length as $\ell_{c}\ll R$, thus a useful dimensionless parameter shall be $\alpha=\ell_{c}/R$. 

In order to solve the above equations (\ref{model1}) and (\ref{model2}), it is convenient to define the marginal density of probability by $\rho(s,t)=\frac{1}{2}\left(P_{+}(s,t)+P_{-}(s,t)\right)$ obtained by integrating out the velocity directions, which represents the density probability to find a particle at the position $s$ at the time $t$, whereas the particle was at certain initial position and initial time. In addition, we introduce the current density of probability by  $\mathbb{J}(s,t)=\frac{v_{0}}{2}\left(P_{+}(s,t)-P_{-}(s,t)\right)$ that take into account the preferential probability direction of the particle motion. From these definitions, it is not difficult to get from  (\ref{model1}) and (\ref{model2}) the 
equations (\ref{ContEq}) and (\ref{CurrentEq}), 
which clearly correspond to a continuity equation, and an exact current equation which gives the actual dynamics of the particle. Remark that (\ref{CurrentEq}) can be re-cast as 
\begin{eqnarray}
\frac{\partial}{\partial t}\left(e^{\frac{t}{\tau_{c}}}\mathbb{J}(s,t)\right)=-\frac{\partial}{\partial s}\left(v_{0}^{2}e^{\frac{t}{\tau_{c}}}\rho(s, t)\right),\label{ContEqAlt}
\end{eqnarray}
which represent a continuity equation for the conjugate quantities $\tilde{\rho}(s,t)=e^{\frac{t}{\tau_{c}}}\mathbb{J}(s,t)$ and $\tilde{J}(s,t)=v_{0}^{2}e^{\frac{t}{\tau_{c}}}\rho(s, t)$.  Now, the above equations, (\ref{ContEq}) and (\ref{CurrentEq}), can be easily decoupled in a one-dimensional telegraphers equations \footnote{ This equation has been used frequently in relativistic diffusion theory, whose applications appear in different contexts ranging from high energy collisions experiments, diffusion of light through turbid media, among others (see \cite{DUNKEL20091, WeissPhysicaA2002} and the references therein)} on the curve $\gamma$
\begin{eqnarray}
\frac{\partial^2}{\partial t^2}\rho(s,t)+\frac{1}{\tau_{c}}\frac{\partial }{\partial t}\rho(s,t)=v_{0}^2 \frac{\partial^2}{\partial s^2}\rho(s,t),\label{tel1}
\end{eqnarray}
where $\partial^2/\partial s^2$ represents the one dimensional Laplacian associated to the one-dimensional space, where $s$ is the arc-length of the curve $\gamma$. Note that $P_{\pm}(s,y)$ and $\mathbb{J}(s,t)$ also satisfy similar telegraphers equations. Strictly speaking $\rho(s,t)$ and $\mathbb{J}(s, t)$   depend on the initial position $s^{\prime}$, and the initial time $t=0$. The initial conditions for the probability density function correspond to $\lim_{t\to 0}\rho(s,  t)=\delta(s-s^{\prime})/L$, which express the fact that at time $t=0$ the particle was at the position $s^{\prime}$; and $\lim_{t\to 0}\partial \rho(s,  t)/\partial t=0$, which express that no particles are introduced at the initial time. Since we unknown the initial conditions for the current probability density, we proceed to solve for $\rho(s,t)$ and then use equations (\ref{ContEq}) and (\ref{CurrentEq}) in order to find a solution for $\mathbb{J}\left(s,t\right)$. 

The one-dimensional space $\gamma$ can be described  through a parametrization  ${\bf X}:I\subset \mathbb{R}\to\mathbb{R}^{2}$ embbedded in a two dimensional Euclidean space. 
Although, the following analysis can be carry out  for any planar Jordan curve,  in the following, we focus in a circular curve $S^{1}$ in order to adapted  the theoretical predictions to the specific experimental conditions. The circle  is described by the parametrization ${\bf X }(s):=R\left(\cos\theta, \sin\theta\right)$,  where $\theta\in\left[-\pi, \pi\right]$, $R$ is the radius of  $S^{1}$, and the arc-length is $s=R\theta$.  In this case, the Laplacian can be written simply as $\frac{1}{R^2}\frac{\partial^{2}}{\partial \theta^{2}}$, whose eigenfunctions are in the set  $\{e^{i m\theta}:m\in\mathbb{Z}\}$ and their eigenvalues are in $\{-\frac{m^2}{R^2}: m\in\mathbb{Z} \}$. The orthonormal relation is given by $\int_{I} d\theta e^{i(m-n) \theta}=2\pi\delta_{mn}$ and the completeness relation in this case corresponds to $\delta(\theta-\theta^{\prime})=\sum_{m\in\mathbb{Z}}e^{im\left(\theta-\theta^{\prime}\right)}
$. In the following, we choose that at the initial time the active particle is at $\theta^{\prime}=0$. 

\subsection{Probability density function $\rho(s,t)$ for the active motion on $S^{1}$}

Here, we present the solution of the telegraphers equation (\ref{tel1}) 
with the appropiate initial conditions. 
 The probability density function, $\rho(s,t)$ is  written as a linear combination of the circular Laplacian eigenfunctions $\sum_{m\in\mathbb{Z}}\tilde{\rho}_{m}(t)e^{im\theta}$, where we identify $\tilde{\rho}_{m}\left(t\right)$ with the {\it Intermediate Scattering Function} defined by 
 \begin{eqnarray}
\tilde{\rho}_{m}\left(t\right)=\int_{I}ds~ e^{-i m\theta}\rho\left(s, t\right),
\end{eqnarray}
 for a particle confined to move on the circle $S^{1}$.  The coefficients $\rho_{m}\left(t\right)$ satisfy the second order differential equation 
$\frac{d^2}{dt^2}\tilde{\rho}_{m}\left(t\right)+\tau^{-1}_{c}\frac{d}{dt}\tilde{\rho}_{m}\left( t\right)+m^2\omega^{2}\tilde{\rho}_{m}\left( t\right)=0$, where it is convenient to define the frequency $\omega=v_{0}/R$, that allows us to write the dimensionless parameter $\alpha=\frac{\ell_{c}}{R}$ also as $\alpha= \omega\tau_{c}$. Thus it is not difficult to show that solutions of last differential equation is a linear combination of $Ae^{-\frac{t}{\tau_{c}}\left(1-\sigma_{m}\right)}+Be^{-\frac{t}{\tau_{c}}\left(1+\sigma_{m}\right)}$, where $\sigma_{m}=\sqrt{1-4m^2\alpha^{2}}$, whose factors $A,B$ can be obtained after imposing the initial conditions in each case. For the case of $\rho(\theta,t)$ the coefficients turns out to be $\tilde{\rho}_{m}\left(0\right)=1/(2\pi R)$, and imposing $\frac{d}{dt}\tilde{\rho}_{m}(0)=0$. After a straightforward calculation, the ISF and the PDF are given by 
\begin{eqnarray}
\tilde{\rho}_{m}\left(t\right)&=&G\left(\frac{t}{2\tau_{c}}, 4m^2\alpha^2\right), \label{ISF}\\
\rho\left(s, t\right)&=&\frac{1}{2\pi R}\left[1+2\sum_{m=1}^{\infty}\cos\left(m\theta\right)G\left(\frac{t}{2\tau_{c}}, 4m^2\alpha^2\right)
\right].\nonumber\\
\label{PDF}
\end{eqnarray}
respectively,  where the function $G(v, w)$ is given by 
\begin{eqnarray}
G\left(v, w\right)=e^{-v}\left[\cosh\left(v\sqrt{1-w}\right)+\frac{\sinh\left(v\sqrt{1-w}\right)}{\sqrt{1-w}}\right].\nonumber\\
\end{eqnarray}
It is noteworthy to mention that the probability density function  is given through a cosine Fourier so that it is symmetric under the interchange $\theta\to-\theta$.  The orthonormal basis in this case is given by $\{1/\sqrt{2\pi R}, \cos\left(m\theta\right)/\sqrt{\pi R}\}$, and its completeness relation is given by (\ref{id1}). This Fourier series (\ref{PDF}) is normalized with the perimeter of the circle, i.e $\int_{I}ds ~\rho (s, t)=1$, where the line element $ds=Rd\theta$ and $I=\left[-\pi, \pi\right]$.

\begin{figure}[ht]
    \centering
    \includegraphics[scale=0.55]{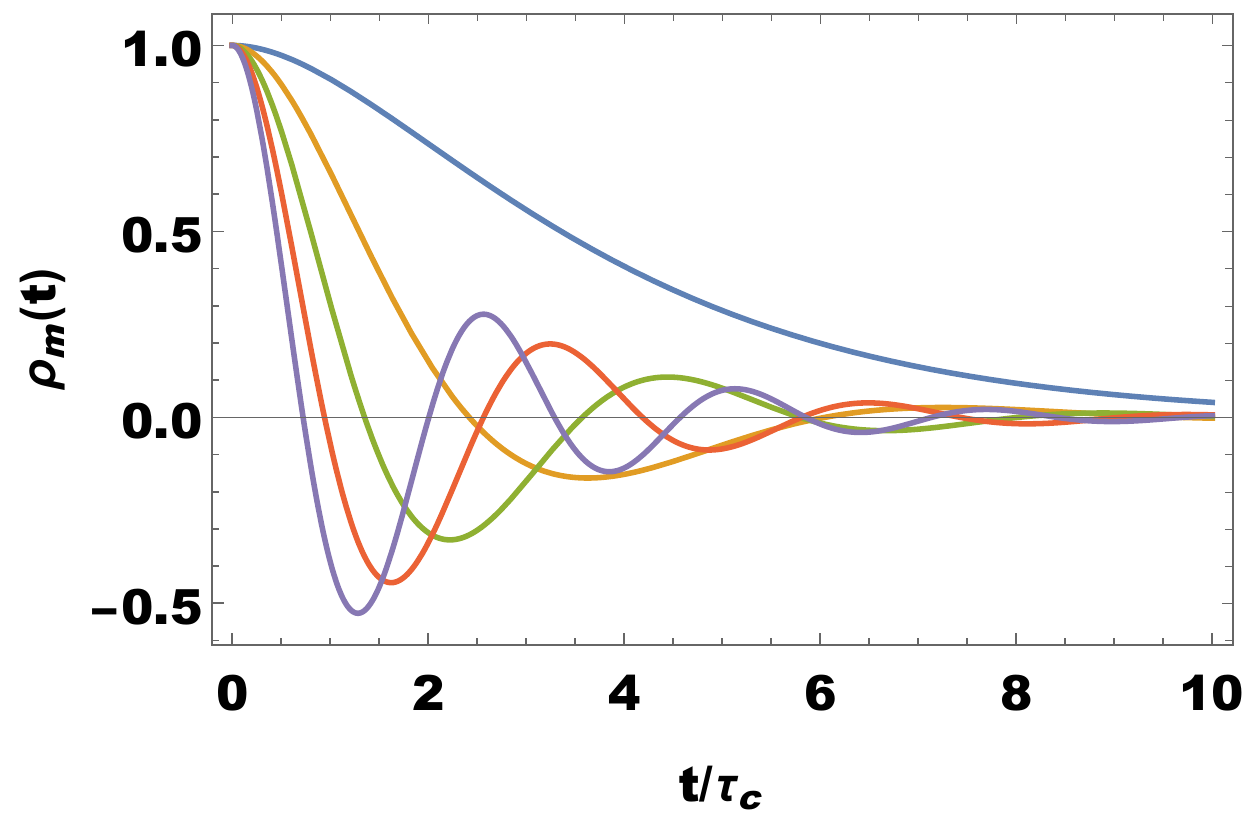}\\
    \includegraphics[scale=0.55]{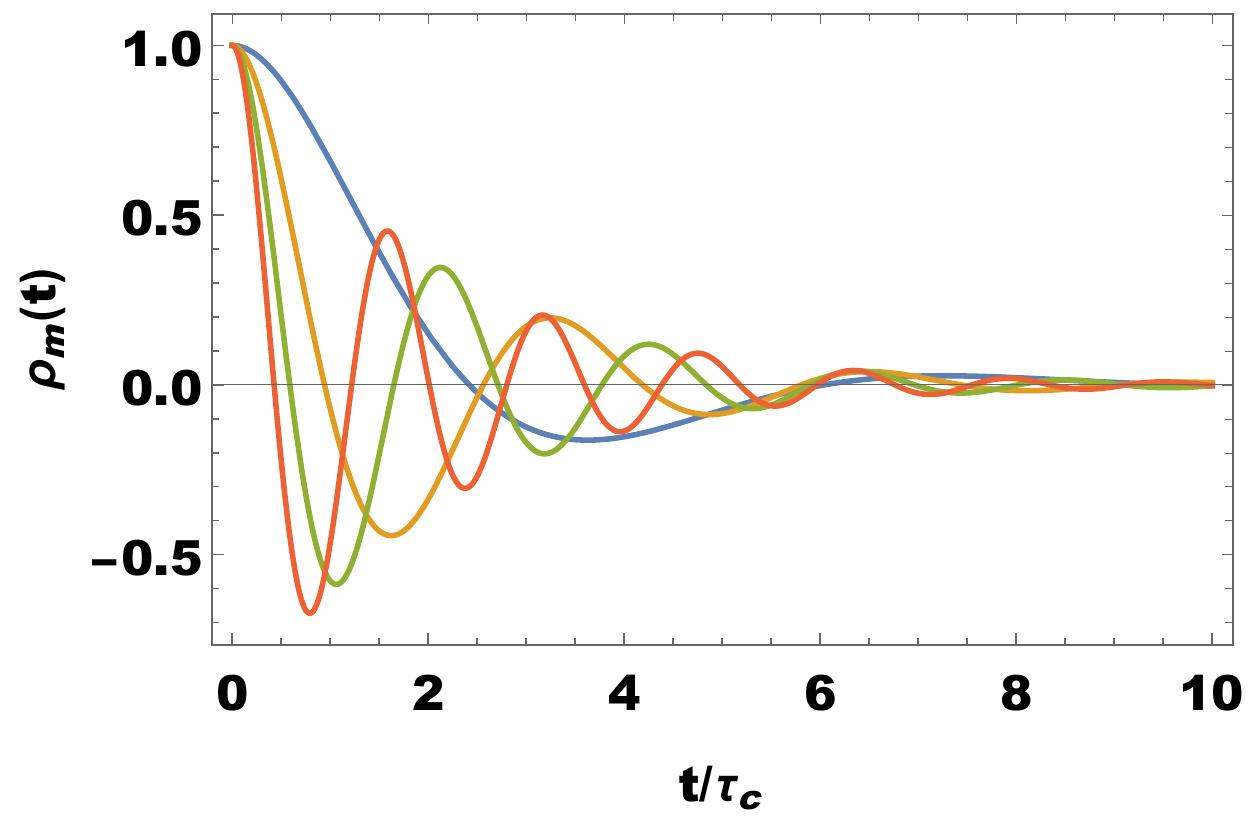}
    \caption{Intermediate scattering function (ISF) (Eq. \ref{ISF}) from the theory of Run and Tumble. Top: ISF for $m=1$ (Eq. (\ref{modomon})), and for $m=2,3,4,5$ (Eq. \ref{ISF}) with $\alpha< 1/2$. Down: ISF for $m=1, 2,3,4,5$ and $\alpha>1/2$.}
    \label{fig1}
\end{figure}

In Fig. (\ref{fig1}), we explore the behavior of the ISF (\ref{ISF}) versus time scaled with the persistent time $\tau_{c}$. Let us note in (\ref{ISF}) that for each $m$ there is a value for $\alpha$ where the intermediate scattering function  transforms its behavior from a monotonous to a oscillating behavior. This value can be determined by making $\sigma_{m}=0$, that occurs when $\alpha$ acquires the values $1/(2m)$ for $m=1, 2, \cdots$. In particular, for $m=1$ let us call $\alpha^{*}:=1/2$, thus one can notice that ISF has an oscillating behavior  for $\alpha>\alpha^{*}$ for each $m\in \{1,2, \cdots\}$; whereas for $\alpha\leq\alpha^{*}$ the ISF function has a monotonous behavior just for the mode $m=1$, and a oscillating one for the rest of the modes with $m>1$. This analysis implies that for $\alpha>\alpha^{*}$ all the terms involved in the series of the probability density (\ref{PDF}) have an oscillating behavior, whereas for   $\alpha\leq\alpha^{*}$ the first term, with $m=1$, is monotonous while the remaining terms, for $m>1$,   have  oscillating behavior. In particular, the ISF for the mode $m=1$ at exactly $\alpha=\alpha^{*}$ is 
\begin{eqnarray}
\left.\tilde{\rho}_{m=1}(t)\right|_{\alpha=\alpha^{*}}=e^{-\frac{t}{2\tau_{c}}}\left(1+\frac{t}{2\tau_{c}}\right).\label{modomon}
\end{eqnarray}

The previous analysis shown  in Fig. (\ref{fig1}) of the ISF (\ref{ISF}), allows us to consider three  dynamical states depending on the parameter $\alpha$, namely, the {\it kinematic state} (KS), when $\alpha\gg 1$;  the {\it diffusive state} (DS), when $\alpha\ll 1$; and the {\it active state} (AS) in between the former and latter. Recall that $\alpha=\ell_{c}/R$, thus KS means a large persistent length, wheres DS a small persistent length. For the active state,  we need to consider the full expression of the PDF (\ref{PDF}).  For the kinematic state, it is convenient to write $\tau_{c}=\alpha/\omega$, thus  the leading  term of the $G$ function is  $G\left(\omega t/(2\alpha), 4m^{2}\alpha^2\right)\simeq \cos\left(m\omega t\right)$, where  the frequency $\omega=v_{0}/R$ is kept fixed. Now, after substituting in (\ref{PDF}) and using (\ref{id1}) one has that the probability density function in the KS corresponds to 
\begin{eqnarray}
\rho_{k}\left(s, t\right)=\frac{\delta\left(\theta-\omega t\right)}{ R}.\label{PDFkinem}
\end{eqnarray}
In the kinematic state, this PDF represents a sharp pulse moving around the circle with uniform angular velocity  $\omega=v_{0}/R$.

Now, for the diffusive regime, {\it i.e} $\alpha\ll 1$, it is not difficult to show that asymptotic expression of $G$ is given by $G\left(\omega t/(2\alpha), 4m^{2}\alpha^2\right)\simeq e^{-m^2 D_{\rm eff} t/R^2}$, as long as the quantity $D_{\rm eff}:=v_{0}^{2}\tau_{c}$ is kept fixed.   Now, after substituting in (\ref{PDF})  one has that the probability density function in the DS corresponds to 
\begin{eqnarray}
\rho_{d}\left(s, t\right)=\frac{1}{2\pi R}\left[1+2\sum_{m=1}^{\infty}e^{-m^2 D_{\rm eff} t/R^2}\cos\left(m\theta\right)
\right].\nonumber\\
\label{PDFdiff}
\end{eqnarray}
This probability density function corresponds to the PDF of a Brownian particle confined in a circle $S^{1}$ \cite{CastroJChemPhys2014}. Now it makes sense to call $D_{\rm eff}$  effective diffusion coefficient.


\subsection{Current probability density $\mathbb{J}(s,t)$}

Here, we present a solution for the current probability density $\mathbb{J}(s,t)$. Since we unknown the initial conditions of the current we proceed to find an expression of the current using the solution for the probability density (\ref{PDF}),  and one of the equations (\ref{ContEq}) and (\ref{CurrentEq}). Following the development  at the appendix (\ref{Appendix C}), the basic procedure consist of integrating (\ref{ContEq}) with respect to $s$ and $t$, respectively. In addition, we choose that at initial time $t=0$, the particle is moving anti-clockwise. Thus the expression for the current probability density is given by 
\begin{eqnarray}
\mathbb{J}\left(s,t\right)&=&\frac{2\omega^2\tau_{c}}{\pi}\sum_{m=1}^{\infty} m\sin \left(m\theta\right) F\left(\frac{t}{2\tau_{c}},\sqrt{1-4m^{2}\alpha^{2}}\right)\nonumber\\
&+&\frac{\omega}{2\pi}e^{-\frac{t}{\tau_{c}}},\label{current}
\end{eqnarray}
where the function $F(v, w)$ is given by
\begin{eqnarray}
F\left(v, w\right)=e^{-v}\frac{\sinh\left(v\sqrt{1-w}\right)}{\sqrt{1-w}}.
\end{eqnarray}
For the active state (AS) we need to consider the full expression (\ref{current}) for the current probability. 

Now, in the following we determine the behavior of the current for the kinematic and diffusive states, respectively. In the KS $\alpha\gg \alpha^{*}$, one has that the leading term of $F(\omega t/\alpha, 4m^{2}\alpha)\simeq \sin\left(\omega t\right)/(2m \alpha)$. Now, after substituting in (\ref{current}) and using (\ref{id2}) one has that the probability density function corresponds to 
\begin{eqnarray}
\mathbb{J}_{k}\left(s, t\right)=\omega\delta\left(\theta-\omega t\right),\label{Currentkinem}
\end{eqnarray}
which using (\ref{PDFkinem}) it can be written as $\mathbb{J}_{k}\left(s, t\right)=v_{0}\rho_{k}(s, t)$ expressing the current of  the sharp pulse moving with constante velocity $v_{0}$ around the circle. 

Now, for the DS, {\it i.e} $\alpha\ll \alpha^{*}$, it is not difficult to show that the asymptotic expression for $F$ is given by $F(\omega t/\alpha, 4m^{2}\alpha)\simeq \frac{1}{2}e^{-D_{\rm eff}t m^{2}/R^2}$ as long as $D_{\rm eff}$ kept is fixed. Now, after substituting in (\ref{current}) the current has the asymptotic behavior
\begin{eqnarray}
\mathbb{J}_{d}(s,t)=\frac{D_{\rm eff}}{\pi R^2}\sum_{m=1}^{\infty} m\sin \left(m\theta\right)e^{-\frac{D_{\rm eff}t m^{2}}{R^2}}
\end{eqnarray}
Notice, as expected,  that in this state is satisfies  the transport Fick's law $\mathbb{J}_{d}(s,t)=-D_{\rm eff}\frac{\partial}{\partial s}\rho_{d}(s,t)$.

\section{Expectation values of observables} 

In this section, we determine the expectation values of several physical observables using the equation (\ref{PDF}) in order to contrast with the experiment described above. In particular, we focus in the following quantities, namely, $\hat{v}$, the direction velocity; the normal vector ${\bf N}=\left(\cos\theta, \sin\theta\right)$; the tangent vector ${\bf T}=(-\sin\theta, \cos\theta)$; the Euclidean displacement $\Delta{\bf R}:={\bf X}(s)-{\bf X}(s^{\prime})$, and the angular displacement $\Delta s=R(\theta-\theta^{\prime})$.  In what follows, we study the behaviour of the mean-values and mean-squared values of these quantities for the three behaviours KS, IS and DS. 

\subsubsection{Mean-value of direction velocity $\hat{v}$}

The stochastic variable $\hat{v}$ gives the direction to the right $(+)$ or to the left $(-)$ of the particle on each point on the circle. The mean-value $\left<\hat{v}\right>$ can be computed using the current probability $\mathbb{J}\left(s, t\right)$ as follows
\begin{eqnarray}
\left<\hat{v}\left(t\right)\right>=\frac{1}{v_{0}}\int_{I}ds~\mathbb{J}\left(s,t\right)
\end{eqnarray}
Now, by direct calculation using (\ref{current}) one has
\begin{eqnarray}
\left<\hat{v}\left(t\right)\right>=\left\{
\begin{array}{cc}
0, &  \alpha\ll 1,\\
\\
e^{-t/\tau_{c}}, &  0<\alpha<\infty, \\
\\
1, &  \alpha\gg 1.
\end{array}
\right.
\end{eqnarray}
Notice that at the diffusive state ($\alpha\ll 1$) the active particle motion does not have any preferential velocity direction. In the kinematic state  the average direction is exactly $\left<\hat{v}\right>=1$, consistent with the leading expression (\ref{Currentkinem}), meaning that in the KS the particle moves anti-clockwise for all the time, whereas in the active state the particle in average moves also  anti-clockwise, however, it damped out exponentially as the time is increased. 

\subsubsection{Mean-value of the geometrical quantities ${\bf N}$ and ${\bf T}$}

In the case of the circle, the normal vector is also related to the vector position ${\bf N}={\bf X}(s)/R$ and the tangent vector ${\bf T}$ to the direction of the active particle vector velocity.  Now, by direct calculation using (\ref{PDF}), (\ref{PDFkinem}) and (\ref{PDFdiff}) the expectation value of the normal vector is
\begin{eqnarray}
\left<{\bf N}\left(t\right)\right>=\left\{
\begin{array}{cc}
e^{-D_{\rm eff}t/R^2}\hat{\bf x},  & ~  \alpha\ll 1,\\
\\
G\left(t/2\tau_{2}, 4\alpha^2 \right)\hat{\bf x},    & ~0<\alpha<\infty,\\
\\
\left(\cos\omega t, \sin \omega t\right),  & ~\alpha\gg 1,
\end{array}
\right.\label{meanvalueN}
\end{eqnarray}
where $\hat{\bf x}=(1,0)$. 
Similarly, the expectation value of the tangent vector is $\left<{\bf T}\left(t\right)\right>=\mathcal{R}_{\pi/2}\left<{\bf N}(s)\right>$, where  $\mathcal{R}_{\pi/2}$ is a ninety degrees anti-clockwise rotation. Notice that no $y-$component appears in $\left<{\bf N}\left(t\right)\right>$ for the DS and IS, since for these states there is not a preferential direction from the initial position, namely, the particle can move with the same probability to the right $(+)$ and to the left $(-)$, while for the KS the particle undergoes a precise uniform circular motion with frequency given by $\omega$. 


\subsubsection{Mean-value and mean-squared value of Euclidean displacement}

\begin{figure}[ht]
    \centering
    \includegraphics[scale=0.35]{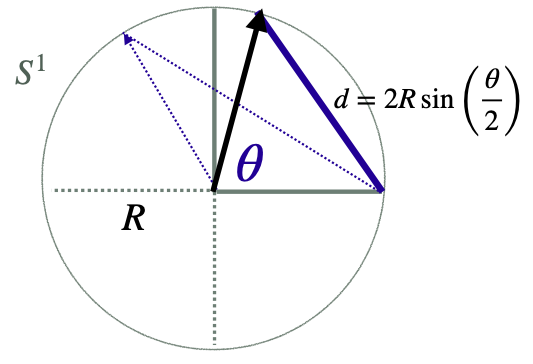}
    \caption{Schematic image of a circle with a chord of length $d$. }
    \label{fig2}
\end{figure}

The Euclidean displacement $\Delta{\bf R}$ can be thought as a {\it fake} physical displacement, because  the active particle confined to the circle $S^{1}$ does not displace along the vector defined by $\Delta{\bf R}$. However, the Euclidean displacement is an observable measurable from experimental point of view that captures the stochastic motion of the particle.  The expectation value can be written in terms of $\left<{\bf N}\left(t\right)\right>$ (\ref{meanvalueN}) as $\left<\Delta{\bf R}\left(t\right)\right>=-R\left(\hat{\bf x}-\left<{\bf N}\left(t\right)\right>\right)$. For the mean-square Euclidean displacement (MSED) $\Delta{\bf R}$ one has $(\Delta {\bf R})^2=2R^2\left(1-\cos\theta\right)$, according to its definition, thus it is enough to compute the expectation $\left<\cos\theta\left(t\right)\right>$, but this quantity is given by $\left<\cos\theta\left(t\right)\right>=G\left(t/2\tau_{c}, 4\alpha^2\right)$ as a consequence of the orthogonality of the terms in the Fourier series (\ref{PDF}). Thus, the MSED is given by 
\begin{widetext}
\begin{eqnarray}
\frac{\left<\Delta{\bf R}^2\left(t\right)\right>}{2R^2}=\left\{
\begin{array}{cc}
 1-e^{-\frac{D_{\rm eff}t}{R^2}},    & ~~~~ \alpha\ll 1,\\
 \\
1-e^{-\frac{t}{2\tau_{c}}}\left[\cosh\left(\frac{t}{2\tau_{c}}\sqrt{1-4\alpha^2}\right)+\frac{\sinh\left(\frac{t}{2\tau_{c}}\sqrt{1-4\alpha^2}\right)}{\sqrt{1-4\alpha^2}}\right],     & ~~~~ 0<\alpha<\infty,\\
\\
2\sin^{2}\left(\frac{\omega t}{2}\right)     &~~~~ \alpha\gg 1,
\end{array} \right.\label{MSDeuclidean}
\end{eqnarray}
\end{widetext}
This equation captures information of the stochastic motion of the particle confined to move on the circle. In the diffusive state ($\alpha\ll 1$) the MSED has the typical behaviour of a Brownian motion on a compact curved manifold, whereas for the kinematic state  the square root of MSD (\ref{MSDeuclidean})    
can be interpreted as the average length  $d$ of the chord shown in the Fig. (\ref{fig2}),  consistent with uniform circular motion exhibited in this regime. 
Furthermore, in the active state, $0<\alpha<\infty$, let us note that similar to the ISF (\ref{ISF}) that there is  the value for $\alpha_{*}=1/2$ where the mean-squared Euclidean displacement transforms its behaviour from a monotonous to a oscillating behaviour. Thus one can notice that MSED has an oscillating behaviour  for $\alpha>\alpha^{*}$, and a monotonous one for $\alpha<\alpha^{*}$. The specific behaviour of the MSED, monotonic or oscillaring, give us  a  signature of a specific active particle that has an intrinsic value of persistence length $\ell_{c}$. In both, diffusive and active states the MSED reachs the limit value of $\left<\Delta{\bf R}^2(t)\right>=2R^2$; this value is known as the geometric limit when there is a uniform probability density in each point of the circle \cite{CastroJChemPhys2014, VilladaCastro2021}.  

\subsubsection{Mean-value and mean-squared value of angular displacement}

The angular displacement $\Delta s=R\theta$ can be thought as the {\it true} physical displacement, because  the active particle indeed moves on the circle $S^{1}$. Also, the angular displacement is another observable measurable from experimental point of view that captures the stochastic motion of the particle.  The expectation value is given by $\left<\Delta s\left(t\right)\right>=0$, which is expected since probability density function is symmetric  (\ref{PDF}).

For the mean-squared geodesic displacement (MSGD), $\left<\Delta s^2\left(t\right)\right>$, we need to carry out a integration by parts $\int_{I}\theta^{2}\cos\left(m\theta\right)d\theta=2(-1)^{m}/m^2$. By a straightforward calculation the MSGD $\left<\Delta s^2\left(t\right)\right>$ is given by 
\begin{eqnarray}
\frac{\left<\Delta s^2\left(t\right)\right>}{R^2}=\frac{\pi^2}{3}
+4\sum_{m=1}^{\infty}\frac{(-1)^{m}}{m^2}\Phi_{m}\left(t, \alpha\right)\label{MSDangular}
\end{eqnarray}
where 
\begin{eqnarray}
\Phi_{m}\left(t, \alpha\right):=\left\{\begin{array}{cc}
e^{-m^2\frac{D_{\rm eff}t}{R^2}}, & \alpha\ll 1,\\
\\
G\left(\frac{t}{2\tau_{c}}, 4m^2\alpha^2\right), & 0<\alpha<\infty.\\
\\
\cos\left(m \omega t\right), & \alpha\gg 1.
\end{array}\right.
\end{eqnarray}

The mean-squared angular displacement for the DS is consistent with previous works \cite{CastroJChemPhys2014}. In addition, the expression for $\left<\Delta s^2\left(t\right)\right>$  in the KS limit is consistent with uniform circular motion shown in the above mean values, since  one can sum up the series in terms of the second Bernoulli polynomial (see Eq. (\ref{id3}) at the appendix), where it can be shown that $\left<\Delta s^2\left(t\right)\right>=R^{2}\left(\omega t\right)^{2}$ for each $t$ in the interval $\left[0, \pi/\omega\right]$, and then repeat the pattern with periods of $2\pi$ for $t>\pi/\omega$. In the active state, similar to the MSED  the mean-squared geodesic displacement transforms its behaviour from a monotonous ($\alpha<\alpha^{*}$) to a oscillating behaviour ($\alpha>\alpha^{*}$).


\section{Experiments vs theoretical predictions
}\label{SectIV}

In this section, we carry out a comparison between the granular magnetic bead experiment described above in section \ref{SectII} and  the theory of Run and Tumble on the circle $S^{1}$ developed in Sect. \ref{SectIII}.  In order to make this comparison, it has been chosen from the theory the mean-squared Euclidean displacement (MSED) $\left<\Delta{\bf R}^{2}(t)\right>$ and mean-squared geodesic displacement (MSGD) $\left<\Delta s^{2}(t)\right>$. Note that from experimental point of view it is measured the MSD $\left<\Delta{\bf r}^{2}(t)\right>$ which is strictly different from the MSED since the later is computed exactly for the motion on the circle $S^{1}$, while MSD captures also radial motions since the channel has finite-size. However, since the thickness of the channel is slightly bigger than the diameter of the particle the following difference $\left<\Delta{\bf R}^{2}(t)\right>-\left<\Delta{\bf r}^{2}(t)\right>$ must be small. Angular displacement can be compute also from the experiment calculating $\theta(t)={\rm arctan}(y(t)/x(t))$, where $x(t)$ and $y(t)$ are extracted from the vector  positions ${\bf r}(t):=(x(t), y(t))$.


\begin{figure}[ht]
\begin{center}
    ~~~~~~  \includegraphics[scale=0.5]{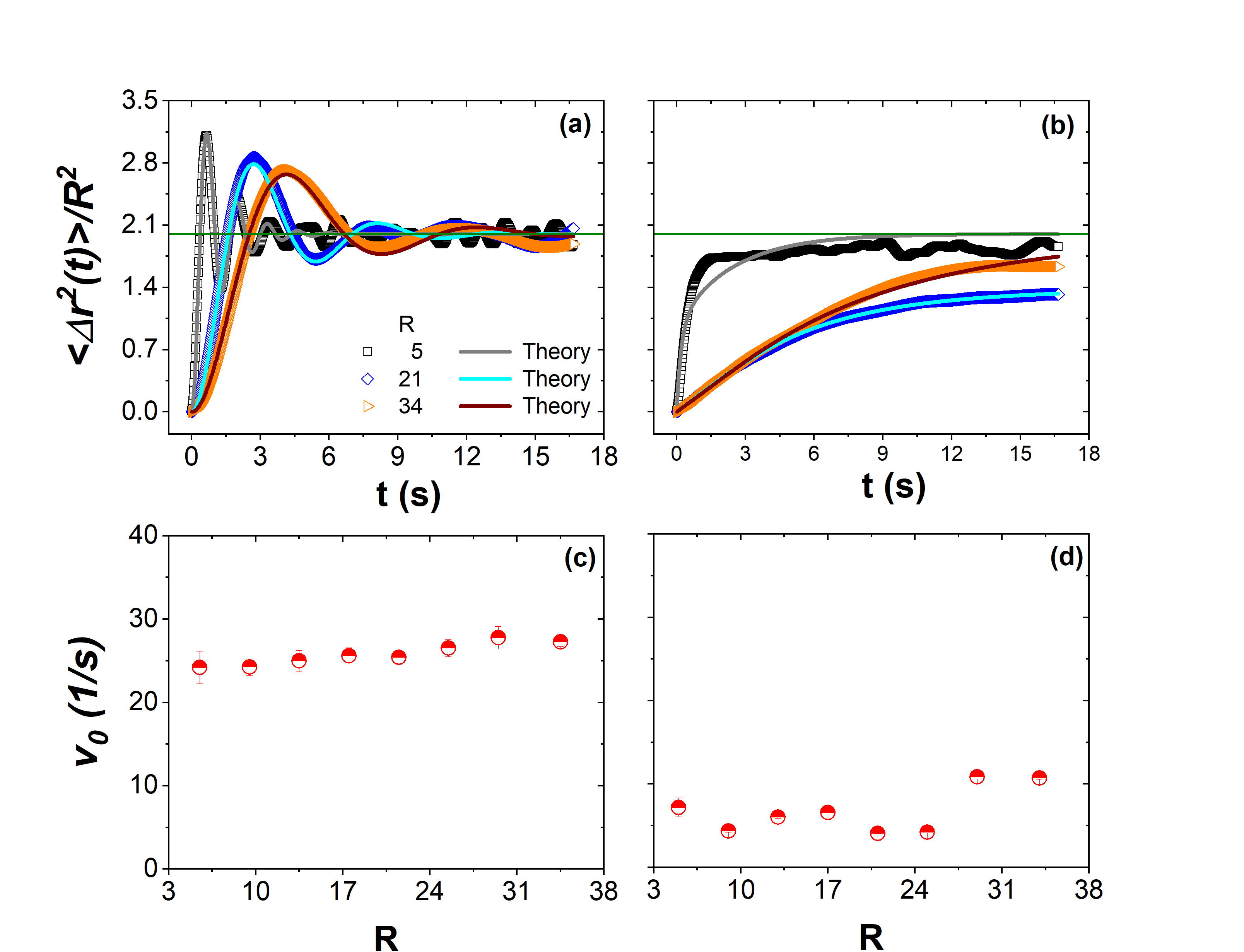}
    \caption{{(Top) Mean-squared Euclidean displacement $\left<\Delta{\bf r}^2\left(t\right)\right>$ for a particle
confined in a circular channel from experiment (open symbols) and Run and Tumble theory (solid lines). Three systems considered with radius  $R/\sigma=5, 21, 34$ at effective temperature $B_{o}=55~{\rm G}$, (a)  for magnetic moment $m_{1}$ (series $S_1$), and (b) for magnetic moment $m_{2}>m_{1}$ (series $S_2$).
The solid lines represent the theoretical prediction (\ref{MSDeuclidean}) adjusting the values of $\alpha$ and $\tau_{c}$ (see the table (\ref{tablaS1S2})). The thin straight line is a reference guide for the eyes, showing  the Euclidean geometrical limit, $2$, value. (Down) Persistence velocity $v_{0}$ versus  radius $R$, (c) for magnetic moment $m_{1}$ (series $S_1$), and (d) for magnetic moment $m_{2}>m_{1}$ (series $S_2$). }}
    \label{7}
\end{center}
\end{figure}


Figure (\ref{7}) (a) shows the comparison between the mean-squared  displacement, $\left<\Delta{\bf r}^2\left(t\right)\right>$,  using experiment series $S_1$, with a particle with the lower magnetization $m_{1}$, and the theoretical predictions encoding in equation $\left<\Delta{\bf R}^2\left(t\right)\right>$ (\ref{MSDeuclidean}). It is observed that the fitted curves very closely reproduce the experimental results. A good experimental and theoretical agreement is also exhibited for the experiment series $S_2$,  with a particle with the higher magnetization $m_{2}$, showed in figure (\ref{7}) (b). In both $S_{1}$ and $S_{2}$ can be observed small discrepancies between the experiment and theory, notoriously observed in the case of small  radius $R=5~\sigma$. These differences are attributed to the finite-size of the channel thickness, which is slightly bigger than the diameter of the particle, implying that the radial degree of freedom of the bead is not completely suppressed.  In both series of experiments, the persistence velocity $v_{0}$ is calculated from a fit of the equation (\ref{MSDeuclidean}) at the active state using the persistence time $\tau_c$ and the dimensionless quantity $\alpha$  as  free parameters through the relation $\alpha=v_{0}\tau_{c}/R$, for each radius $R$ in the range $R/\sigma$ from $5$ to $34$ (see table (\ref{tablaS1S2})).  In particular, it is clear that all values of $\alpha$ in $S_{1}$ are bigger than $1/2$, while in $S_{2}$ are smaller than $1/2$, according to the prediction of the theory.  
Figure (\ref{7}) (c) and (d) show the persistence velocity $v_{0}$ for the experiment series $S_1$, and $S_2$ respectively. It is observed that for the system with magnetic moment $m_{1}$ the particle persistence velocity $v_{0}$ slightly increased as $R$ increases and roughly is a constant with a value around $25.7~ \sigma/{\rm s}$, whereas for the system with magnetic moment $m_{2}$ the persistence velocity has a more abrupt behaviour and the increase is more pronounced, but it has bigger oscillations around the value $5~\sigma/{\rm s}$. Additionally, the magnitude of persistence velocity $v_{0}$ is smaller for the higher magnetic dipole particle than for the smaller magnetic dipole, which is the same trend observed in the experimental results discussed above in relation to the figure (\ref{3}).

\begin{table}[ht]
\caption{In this table we show the values for the parameter $\alpha:=\ell_{c}/R$ and the persistence time $\tau_{c}$ obtained by adjusting equation (\ref{MSDeuclidean}) to the experimental values for the experiment series $S_{1}$ and $S_{2}$.}
\begin{center}
\resizebox{10cm}{!} {
\begin{tabular}{||c|c|c|c|c||}
\hline
\multirow{2}{*}{$R(\sigma)$} & \multicolumn{2}{|c}{$S_1$} &  \multicolumn{2}{c||}{$S_2$} \\
\cline{2-5}
& $\alpha$ & $\tau_c$ (s) &$\alpha$ & $\tau_c$ (s)\\
\hline
5 &2.812 $\pm$	0.042  & 0.581$\pm$	0.009& 0.261	$\pm$ 0.009 & 0.181	$\pm$ 0.013 \\
\hline
9 & 2.325	$\pm$ 0.016 & 0.864$\pm$	0.006 &0.292 $\pm$ 0.001 & 0.604	$\pm$ 0.003\\
\hline
13 &  2.866	 $\pm$ 0.045 & 1.492$\pm$	0.024 & 0.286 $\pm$	0.001 &0.615	$\pm$ 0.003 \\
\hline
14 & 3.182 $\pm$	0.037 & 2.113$\pm$	0.026 & 0.278 $\pm$	0.002 & 0.718 $\pm$	0.004\\
\hline
21 & 1.756 $\pm$	0.013 & 1.451$\pm$	0.013& 0.158 $\pm$	0.001 & 0.813 $\pm$	0.001  \\
\hline
25 & 2.211 $\pm$	0.031 & 2.084$\pm$	0.031& 0.147 $\pm$	0.001 & 0.875 $\pm$	0.001 \\
\hline
29 & 1.586 $\pm$	0.029& 1.655$\pm$	0.035& 0.297 $\pm$	0.002 &	0.795 $\pm$	0.004  \\
\hline
34 & 1.522 $\pm$	0.009& 1.898$\pm$	0.013& 0.288 $\pm$	0.001 &	0.911 $\pm$	0.002\\
\hline
\end{tabular}}
\end{center}
\label{tablaS1S2}
\end{table}


\begin{figure}[ht]
\begin{center}
        \includegraphics[scale=0.5]{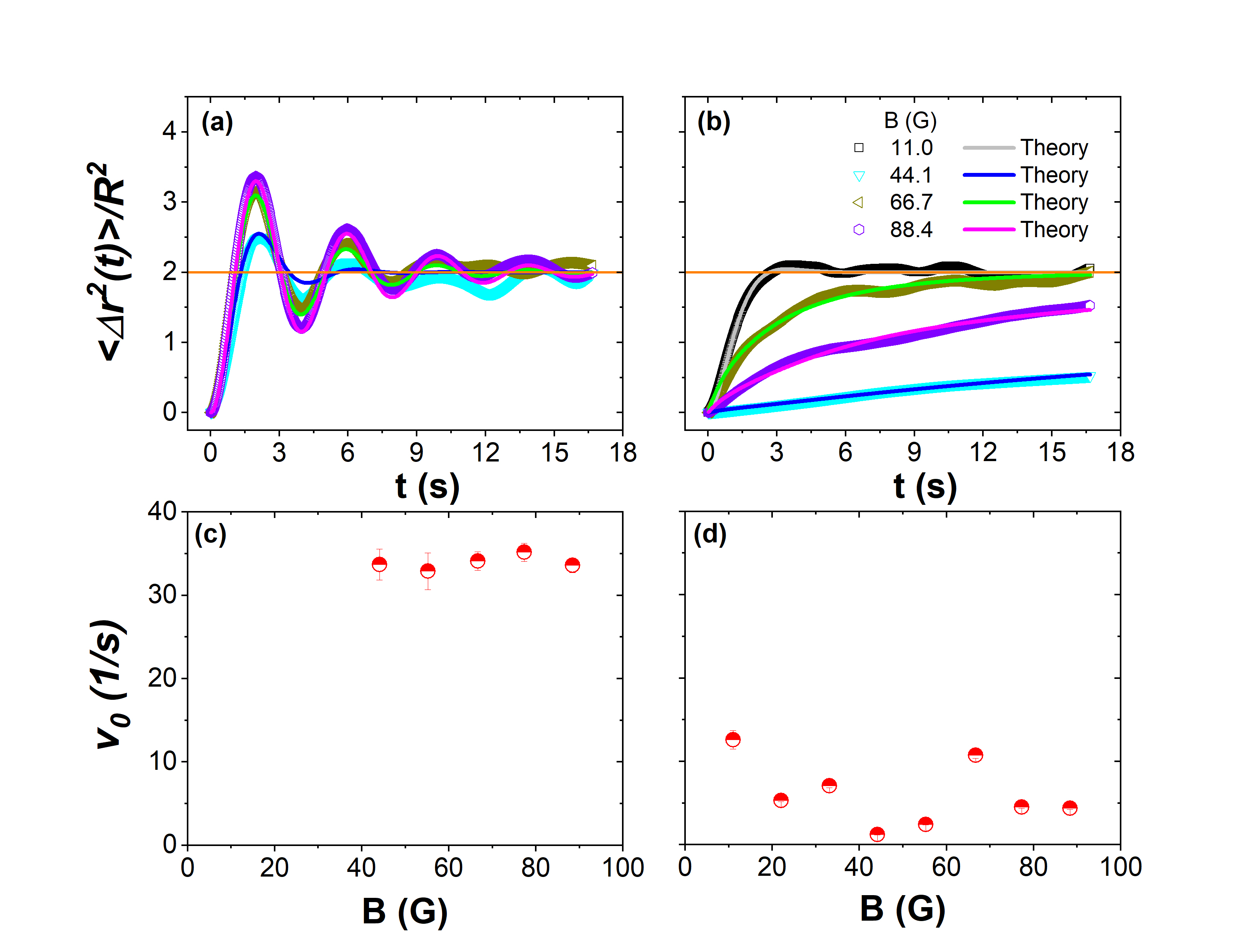}
    \caption{{(Top) Mean-squared Euclidean displacement $\left<\Delta{\bf r}^2\left(t\right)\right>$ for a particle
confined in a circular channel from experiment (open symbols) and Run and Tumble theory (solid lines). Three systems considered with magnetic field  $B=11 ~{\rm G}$ to $B=88.4 ~{\rm G}$, (a)  for magnetic moment $m_{1}$ (series $S_3$) with radius $R=20~\sigma$, and (b) for magnetic moment $m_{2}>m_{1}$ (series $S_4$) with radius $R=21~\sigma$.
The solid lines represent the theoretical prediction (\ref{MSDeuclidean}) adjusting the values of $\alpha$ and $\tau_{c}$ (see the table (\ref{cristal24})). The thin straight line is a reference guide for the eyes, showing  the Euclidean geometrical limit, $2$, value. (Down) Persistence velocity $v_{0}$ versus  radius $R$, (c) for magnetic moment $m_{1}$ (series $S_3$), and (d) for magnetic moment $m_{2}>m_{1}$ (series $S_4$).
}}
   \label{8}
\end{center}
\end{figure}


\begin{table}[ht]
\caption{In this table we show the values for the parameter $\alpha:=\ell_{c}/R$ and the persistence time $\tau_{c}$ obtained by adjusting equation (\ref{MSDeuclidean}) to the experimental values for the experiment series $S_{3}$ and $S_{4}$. Notice that values $B=11, 22, 33~{\rm G}$ are absent in series $S_{3}$ since the reasons mentioned at (\ref{SectII}). }
\begin{center}
\resizebox{10cm}{!} {\begin{tabular}{||c|c|c|c|c||}
\hline
\multirow{2}{*}{$B({\rm G})$} & \multicolumn{2}{|c}{$S_3$} &  \multicolumn{2}{c||}{$S_4$} \\
\cline{2-5}
& $\alpha$ & $\tau_c$ (s) &$\alpha$ & $\tau_c$ (s)\\
\hline
11 & & &0.283$\pm$	0.014 & 0.450$\pm$	0.007\\
\hline
22 & && 0.236$\pm$	0.001& 0.888$\pm$	0.001\\
\hline
33 &  && 0.300$\pm$	0.001 & 0.850$\pm$	0.001\\
\hline
44 & 1.328 $\pm$	0.025 & 0.828 $\pm$	0.020 & 0.060$\pm$	0.002 & 0.964$\pm$	0.001\\
\hline
55 & 3.707 $\pm$	0.100 & 2.367$\pm$	0.066 & 0.106$\pm$	0.001 & 0.867$\pm$	0.001\\
\hline
66 & 2.653	$\pm$ 0.027 & 1.634 $\pm$	0.017 & 0.327$\pm$	0.002 & 0.609$\pm$	0.004\\
\hline
77 & 3.932	$\pm$ 0.037 & 2.348	 $\pm$ 0.022 & 0.179$\pm$	0.001 &	0.794$\pm$	0.001 \\
\hline
88 & 3.667	$\pm$ 0.026 & 2.293	$\pm$ 0.017 & 0.179	 $\pm$0.001&0.822$\pm$	0.001	\\
\hline
 \end{tabular}}
\end{center}
\label{cristal24}
\end{table}

Figure (\ref{8}) (a) shows  the mean-squared displacement for the particle with lower magnetization $m_{1}$ in a circle with radius $R=20~\sigma$, and figure (\ref{8}) (b) for the particle with the higher magnetization $m_{2}$ in a circle with radius $R=21~\sigma$, for the experiment series $S_3$, and $S_4$, respectively. In both series it has been varied the magnitude of the amplitude of the magnetic field $B_{o}$ from $11~{\rm G}$ to $88.4~{\rm G}$. Also, both figures show the comparison with the theoretical expression (\ref{MSDeuclidean}). It is observed that the fit between experiments and theory has a very good agreement. The persistence velocities $v_{0}$ were calculated from the values obtained with the fits (see table (\ref{cristal24})). In particular, it is clear that all values of $\alpha$ in $S_{3}$ are bigger than $1/2$, while in $S_{4}$ are smaller than $1/2$, according to the prediction of the theory.  In these experiment series, also small discrepancies can be observed between the experiment and theory, observed
in the case of small magnetic amplitude $B=11~{\rm G}$ and $B=44~{\rm G}$ (series $S_{3}$). These differences are attributed also to the small radial motion  and because the particle does not absorb enough energy from the magnetic field in order to self-propel along the circle. Figure (\ref{8}) (c) shows that there exists a soft dependence of $v_{0}$ on the effective temperature and roughly is a small oscillation  around $33.9~ \sigma/{\rm s}$, this is clear for the lowest and the highest $B_{o}$ values of the temperature. For the intermediate values, $v_{0}$ slightly increases as $B_{o}$ increases. The physical explanation of the fact that for the highest effective temperature $B_{o}=88\,~{\rm G}$ the value of $v_{0}$ decreases slightly is due to the magnetic interaction between the particle and the magnetic field is higher, so that the particle exhibits more changes in its direction leading to a more erratic motion. This behaviour can also be observed in the experiment series $S_4$, where the particle has higher magnetization. Figure (\ref{8}) (d) shows that persistent velocity $v_{0}$ has a more abrupt behaviour and  bigger oscillations, with a general trend that it decreases as $B_{o}$ increases. At a low magnetic field, the particle travels longer paths, but as the magnetic field increases, the particle frequently changes direction, so the particle trajectory is more random. Thus, certainly, the intensity of the interaction between the field and the particle magnetization has an important influence on the direction and inertial components of the particle motion, namely, it has an impact on the tumbling behaviour that it manifests in the erratic motion of the particle. Note that in both Fig. (\ref{7}) and Fig. (\ref{8}) can be observed the limit value $\left<\Delta {\bf R}^2(t)\right>/R^2=2$ in the geometric regime. 

\begin{figure}[ht]
\begin{center}
    \includegraphics[scale=0.5]{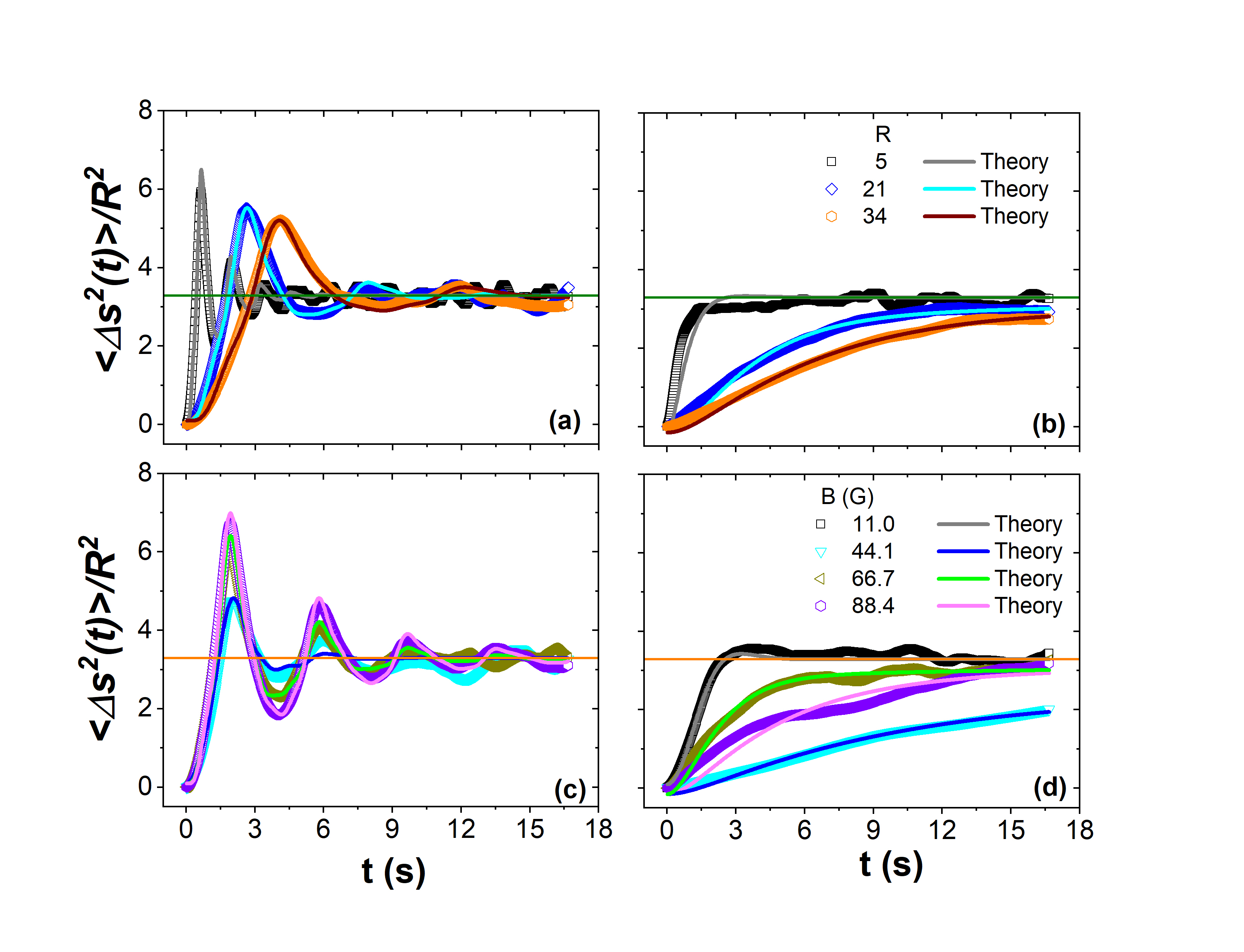}
    \caption{{
     Mean-squared geodesic displacement $\left<\Delta{s}^2\left(t\right)\right>$ for a particle
confined in a circular channel from experiment (open symbols) and Run and Tumble theory (solid lines). Three systems considered with radius  $R/\sigma=5, 21, 34$ at effective temperature $B_{o}=55~{\rm G}$, (a)  for magnetic moment $m_{1}$ (series $S_1$), and (b) for magnetic moment $m_{2}>m_{1}$ (series $S_2$). Additionally, four systems considered with magnetic amplitude $B/{\rm G}=11, 44.1, 66.7, 88.4$, (c) for magnetic moment $m_{1}$ (series $S_{3}$) with radius $R=20~\sigma$, and (d) for magnetic moment $m_{2}>m_{1}$ (series $S_{4}$) with radius $R=21~\sigma$. The solid lines represent the theoretical prediction (\ref{MSDangular}) adjusting the values of $\alpha$. The thin straight line is a reference guide for the eyes, showing  the geodesic geometrical limit, $\pi^2/3$, value. 
    }}
   \label{geodesicos}
\end{center}
\end{figure}

Figure (\ref{geodesicos}) shows the comparison between the theoretical predictions and experimental results using the mean-squared geodesic displacement $\left<\Delta s^2(t)\right>$. It is shown just three cases for the experiment series $S_{1}$, $S_{2}$, and $S_{3}$, and four cases for the experiment series $S_{4}$. In this comparison it is considered only the dimensionless parameter, $\alpha$, as free parameter, whereas the 
persistence time, $\tau_{c}$ corresponds to the value used to adjust  mean-squared Euclidean displacement in the corresponding system. 
The curves of the mean-square geodesic displacement indicate that the fit among experimental data and theoretical prediction (\ref{MSDangular}) has a good agreement. The general trend of the MSGD curves is similar to the corresponding Euclidean displacements cases. In the series $S_1$, figure (\ref{geodesicos}) (a), and $S_3$, figure (c), the maximum of the oscillations are higher than the corresponding euclidean cases. Similarly, series $S_2$, figure (\ref{geodesicos}) (b), and $S_4$, figure (\ref{geodesicos})  (d), show similar behaviour to the corresponding Euclidean cases. Additionally, the notoriously difference between the MSGD and MSED is the geometric regime which in the geodesic case corresponds at the long-time regime to $\left<\Delta s^{2}(t)\right>/R^2= \frac{\pi^2}{3}\approx 3.28$.  

\begin{figure}[ht]
\begin{center}
    \includegraphics[scale=0.5]{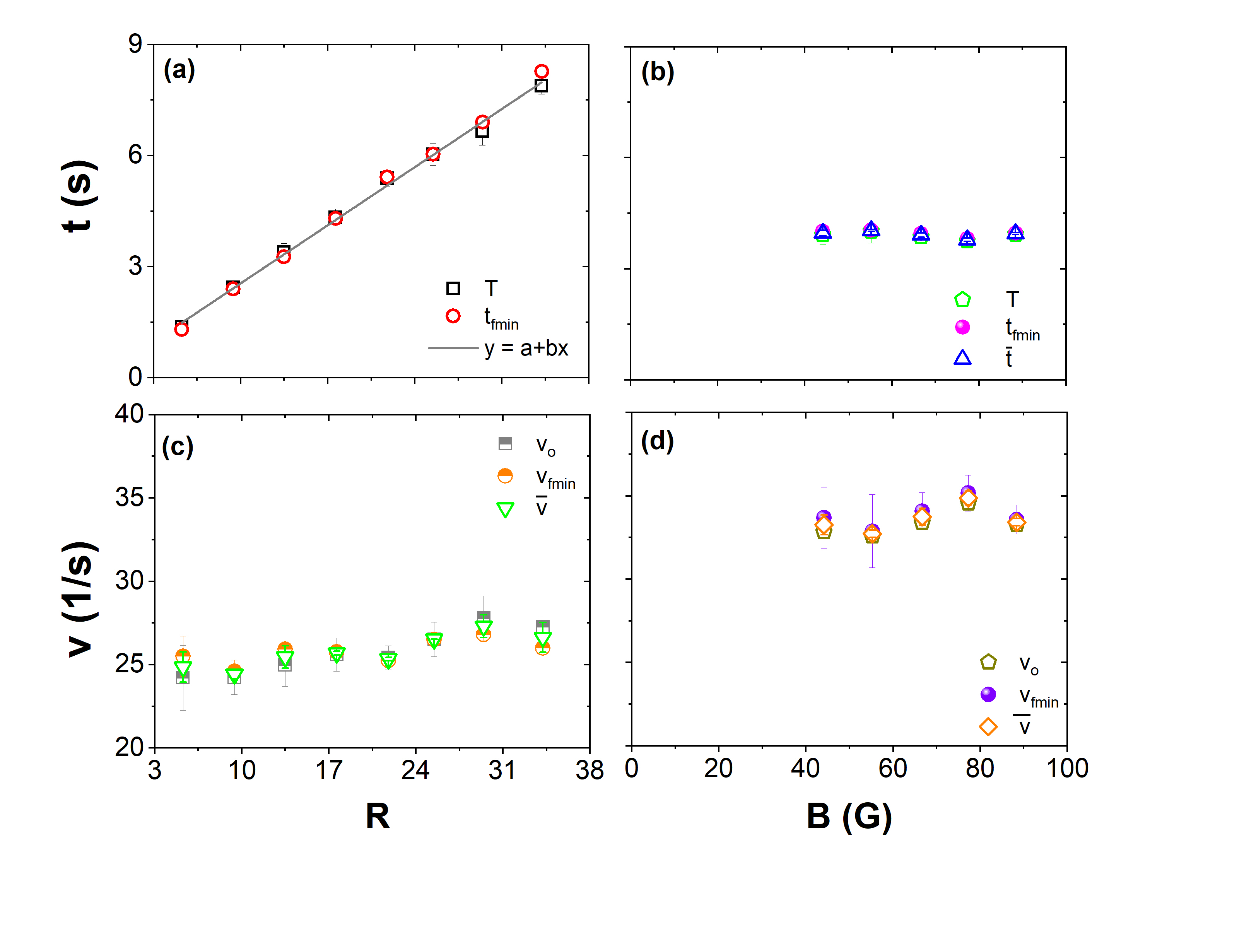}
    \caption{{  Period of oscillation from experiment (open circles symbols) and $T=2\pi R/v_{0}$ where values of $v_{0}$ extracted from Fig. (\ref{7}) as a function of the (a)  radius of the circular channel, $R$, and (b) amplitude of the magnetic field. Persistence velocity $v_{0}$ as a (a) function of the radius  of the circular channel, $R$, and (b) amplitude of the magnetic field $B_{0}$. }}
   \label{11}
\end{center}
\end{figure}

Figures (\ref{11}) (a) and (b) show the comparison between the period, $T$, obtained as the ratio of the perimeter $2\pi R$ and the persistence velocity $v_{0}$ (whose values are extracted from  Fig. (\ref{7}) (c)), and the period obtained directly from the experimental data corresponding to the mean-square displacements curves, measured from the beginning up to the first minimum, $t_{min}$, for the experiment series $S_1$ and $S_3$, respectively. It is observed that the values are very close to each other. Figure (\ref{11}) (a) shows that the period is linear with the radius, implying that persistence velocity for the particles with smaller  magnetic moment  does not depend on the radius of the circle. 
Figure (\ref{11}) (b) shows that the period has only a slight dependence on the effective temperature. Furthermore, a comparison between the persistence velocity obtained from the fitting through the theoretical model (\ref{MSDeuclidean}) and the velocity obtained as the ratio of the perimeter and the period obtained directly from the experiment, it is shown in Figures (\ref{11}) (c) and (d). Again it is observed that in both series $S_{1}$ and $S_{3}$, persistence velocity values are very close to each other.  In Figure (\ref{11}) (c) it is observed that the velocity slightly increases as the radius increases. Figure (\ref{11}) (d) shows that although the slight dependence of the persistence velocity on the effective temperature one observes that $v_{0}$ oscillates around the value $33~\sigma/s$. 
These can be understood as a consequence of an extremely weak and strong magnetic interaction between the field and the particle.


\section{Concluding remarks
}\label{SectV}
 
In this paper, we investigated the random motion of an active particle confined along a circle through a non-vibrated granular experiment contrasting with the model of run and tumble used to describe the stochastic motion of an active particle. On the one hand, the experiment consists of confining a magnetized metallic ball into a circular channel subjected to an alternating magnetic field that causes stochastic motion. In each experiment, the positions of the particle were determined after the data analysis of the recorded videos, which allow calculating from the experiment the mean-squared displacement. On the other hand, we have developed the model of run and tumble to describe the self-propel motion of an active particle characterized by two parameters, namely, the persistence velocity $v_{0}$ and the tumble rate $\lambda=\tau_{c}^{-1}$, where $\tau_{c}$ is the persistence time. Using this model we carried out an exact statistical physics analysis that allow comparison with the experiment, finding a good agreement between the experimental results and the theoretical predictions. 

The theoretical predictions of the run and tumble model on the circle established the existence of a transition between two-state of motion: an erratic motion (or disordered phase),  and persistent motion (or ordered phase). From the viewpoint of the theory, the change of the persistence length,  $\ell_{c}=v_{0}\tau_{c}$, of a particle triggers a transition phase from a disorder to an ordered phase. Particularly,  at the level of the MSD, the ordered phase is characterized by an oscillating function when $\ell_{c}>R/2$, while at the disordered phase its behaviour corresponds to an increasing monotonic function, for $\ell_{c}<R/2$, that saturates to a certain finite value due to the compactness of the circle. Additionally from the theory, one can show that for $\ell_{c}\gg R$ the movement of the particle corresponds to a  uniform circular motion, whereas for $\ell_{c}\ll R$ the movement corresponds to the usual Brownian motion on the circle.  

Now from the experimental point of view, the magnetization $m$ is the only essential property of the particle, and the oscillating behaviour appears for particles with the lower magnetization, while the monotonic behaviour is for the higher value of magnetization. Thus, one can conclude qualitatively that the higher the persistence length of a particle is, the lower the magnetization it is, at least within the experimental limit of validity considered in the present experiment. Particularly, a specific manifestation of the two states of motion predicted by the theory can be observed in the figure (\ref{3}) where  particles with lower magnetization exhibit larger traces whereas the particle with higher magnetization travels a smaller portion of the entire circular channel. Additionally, is has been shown that at the short-time the movement is superdiffusive, almost ballistic, in the order phase, while in the disordered phase, at short-time regime, the motion becomes approximately diffusive. We have shown that this effect can be modulated by varying the particle-field interaction. It is observed that as the interaction between the particle and magnetic field increases, the particle experiments more changes in its direction, going from a superdiffusive to a diffusive behaviour. This interaction can be modulated varying the magnetization of the particle, namely, by exposing it to a static magnetic field of different intensities, or along different exposition times. 

After comparing the theory and experiment, it has been shown that the magnetized metallic balls under the alternating magnetic field have the most salient properties exhibited by the active particle matter systems. The magnetization of the ball corresponds to the intrinsic property in this case, and its interaction with the time-varying magnetic field allows the particle to extract energy to transform it into a self-propel motion along the quasi-$1D$ circular channel. In particular, it has been shown that this granular non-vibrated experiment describe the main characteristics determined by the run-and-tumble model (originally to describe the bacterial motion \cite{MartensEPJE2012}), at least in one-dimensional confinement in absence of exclusion effects.

The present work can be extended in several directions. Using the same experiment setup it remains to carry out an analysis using more than two different values of magnetizations to establish a quantitative law between the persistence length and the magnetization of the ball.  Now that we have proved that the magnetized metallic balls behave as active particles, one can address the single-file diffusion problem of studying the interacting active particle system confined in quasi 1D circular channels, this analysis can be extended the situation already known in colloidal particle systems \cite{VilladaCastro2021}. Changing the experiment set up by replacing the circular channel with a concave surface plate can address the problem of a single active particle moving on a curved surface \cite{CastroPRE2018, ApazaSoftMatter2018J}. For instance, using this type of experiment one can address the phase transition triggering in a  single active particle in a spherical surface as it is predicted theoretically in \cite{CastroPRE2018}. Furthermore, since at $1D$ the telegrapher equation (\ref{tel1}) is common in various models of active particle. It remains the open question to answer what is the most adequate model to describe the stochastic motion of the magnetized active particle in higher dimensions. This could be addressed by considering in a $2D$ situation the Active Brownian Motion \cite{Romanczuk2012, Sevilla2014}, the Run and Tumble model \cite{MartensEPJE2012}, and the Generalized Active Brownian Motion model proposed recently in \cite{SevillaCastro2021}.

\begin{acknowledgments}
The partial financial support by CONACyT, México, through the Grants No. 731759 and A1-S-39909 is acknowledged.
\end{acknowledgments}

\appendix

\section{Run and tumble model in manifolds}
\label{sec:sample:appendix}
In this section, it is introduced the run and tumble model \cite{MartensEPJE2012} for an active particle moving in a $d-$dimensional curved spaces $\mathcal{M}$. A particle follows this model also has an internal degrees of freedom that dictate the direction of motion according to the  condition that the velocity is ${\bf v}=v_{0}\hat{v}$, where $\hat{v}\in S^{d-1}$ and $v_{0}$ is a constant persistence velocity. Thus the phase-space available for this particle corresponds to $\mathcal{M}\times S^{d-1}$.  The model in the curved space is given by 
\begin{eqnarray}
\frac{\partial}{\partial t}P(x, \hat{v}, t)&+&v_{0}\hat{v}\cdot \nabla P(x, \hat{v}, t)=-\lambda P(x, \hat{v}, t)\nonumber\\&+&\lambda \int_{S^{d-1}}\frac{d\hat{v}}{V(S^{d-1})} P(x, \hat{v}, t)\label{RTcurvedspace}
\end{eqnarray}
where $\left\{x^{a}\right\}$ with $a=1, \cdots, d$, represents a set of local coordinates and  $\hat{v}\cdot \nabla f=\frac{1}{\sqrt{g}}\hat{v}^{a}\partial_{a}\left(\sqrt{g}f\right)$, where $g$ is the determinant of the tensor metric $g_{ab}$ that defines the Riemannian geometry of the space $\mathcal{M}$, and $V(S^{d-1})$ is the volume of the sphere $S^{d-1}$. In the above model, $\lambda$ is the tumbling rate that gives in average how many tumbling the particle makes in a unit of time.

In the simplest case, when $d=1$, the sphere $S^{0}$ has only two points $S^{0}=\{-1, 1\}$, that are interpreted as the direction to the left $(-1)$ or the right $(+1)$,  the volume $V(S^{0})=2$, and the integration is given adopting the formal expression  
$\int d\hat{v} f(\hat{v})=f(+1)+f(-1)$.
From this definition, we define the probability density function $\rho(s,t)$ and the current probability function
 \begin{eqnarray}
\rho(s,t)&=&\int_{S^{0}}\frac{d\hat{v}}{V(S^{0})}P(s, \hat{v}, t), \\
\mathbb{J}(s,t)&=&\int_{S^{0}}\frac{d\hat{v}}{V(S^{0})}\hat{v} P(s, \hat{v}, t).
\end{eqnarray}

\section{Calculation of the Current probability density $\mathbb{J}(s,t)$}\label{Appendix C}

Here, we present a calculation for the current probability density $\mathbb{J}(s,t)$. We proceed to find an expression of the current using the solution for the probability density (\ref{PDF}), and the equations (\ref{ContEq}) and (\ref{ContEqAlt}). Let us calculate the partial derivative of $\rho(s, t)$ with respect to time, that is, 
{\small\begin{eqnarray}
\frac{\partial \rho}{\partial t}=-\frac{2\alpha^2 e^{-\frac{t}{2\tau_{c}}}}{\tau_{c}\pi R}\sum_{m=1}^{\infty}\cos\left(m\theta\right)\frac{m^2\sinh\left(\frac{t}{2\tau_{c}}\sqrt{1-4m^2\alpha^2}\right)}{\sqrt{1-4m^2\alpha^2}}.\nonumber\\
\end{eqnarray}}
Now, in virtue  of continuity equation (\ref{ContEq}) we equate last expression to $-\frac{1}{R}\frac{\partial\mathbb{J}}{\partial\theta}$. Afterwards, we integrate out the variable $\theta$ such that
{\small\begin{eqnarray}
\mathbb{J}\left(s,t\right)&=&\frac{2\omega^2 \tau_{c}e^{-\frac{t}{2\tau_{c}}}}{\pi}\sum_{m=1}^{\infty}m\sin(m\theta)\frac{\sinh\left(\frac{t}{2\tau_{c}}\sqrt{1-4m^2\alpha^2}\right)}{\sqrt{1-4m^2\alpha^2}}\nonumber\\&+&\psi(t)\label{C2},
\end{eqnarray}}
where $\psi(t)$ is a time function to be determine. To obtain $\psi(t)$, we observe that the current can be also obtain using  (\ref{ContEqAlt}), thus, we calculate $\partial \rho/\partial s$, namely, 
\begin{eqnarray}
\frac{\partial \rho}{\partial s}=-\frac{1}{\pi R^2}\sum_{m=1}^{\infty}m\sin\left(m\theta\right)G\left(\frac{t}{2\tau_{c}}, 4m^2\alpha^2\right).
\end{eqnarray}
Now we integrate out the time variable $t$ both sides of Eq. (\ref{ContEqAlt}), such that, we get for the current
{\small\begin{eqnarray}
\mathbb{J}\left(s, t\right)&=&\frac{2\omega^2 \tau_{c}e^{-\frac{t}{2\tau_{c}}}}{\pi}\sum_{m=1}^{\infty}m\sin\left(m\theta\right)\frac{\sinh\left(\frac{t}{2\tau_{c}}\sqrt{1-4m^2\alpha^2}\right)}{\sqrt{1-4m^2\alpha^2}}\nonumber\\&+&\varphi(s)e^{-\frac{t}{\tau_{c}}}\label{C4}.
\end{eqnarray}}
Now, comparing both expressions (\ref{C2}) and (\ref{C4}) of the current,
one can conclude that $\varphi(s)=\mathbb{J}_{0}$ is a constant independent of $s$ and $\psi(t)=\mathbb{J}_{0}e^{-\frac{t}{\tau_{c}}}$. The constant $\mathbb{J}_{0}$ is determined noting that the integration of the series term in (\ref{C4}) vanished then $\int_{I}ds \mathbb{J}(s, t)=2\pi R\mathbb{J}_{0}e^{-\frac{t}{\tau_{c}}}$, now observing that $\int_{I}ds \mathbb{J}(s, t)=v_{0}\left<\hat{v}(t)\right>$. Now let us choose that at $t=0$ the direction is such that $\left<\hat{v}(0)\right>=1$, thus one has $v_{0}=(2\pi R)\mathbb{J}_{0}$, thus $\mathbb{J}_{0}=v_{0}/(2\pi R)$. Fnally, after substitute this value we got the desired result (\ref{current}). 

\section{Useful mathematical identities}
The following mathematical identities were useful for the analytical calculations
{\small\begin{eqnarray}
\frac{1}{2\pi R}\left(1+2\sum_{m=1}^{\infty}\cos\left(m\theta\right)\cos(m \theta^{\prime})\right)&=&\frac{1}{R}\delta\left(\theta-\theta^{\prime}\right)\nonumber\\
\label{id1}\\
\frac{1}{2\pi R}\left(1+2\sum_{m=1}^{\infty}\sin\left(m\theta\right)\sin(m \theta^{\prime})\right)&=&\frac{1}{R}\delta\left(\theta-\theta^{\prime}\right)\nonumber\\
\label{id2}\\
\sum_{m=1}^{\infty}\frac{\left(-1\right)^{m}}{m^{2}}\cos\left(m y\right)&=&\pi^{2}B_{2}\left(\frac{y}{2\pi}-\frac{1}{2}\right)\nonumber\\
\label{id3}
\end{eqnarray}}
where $B_{2}(x)=x^2-x+\frac{1}{6}$ is the second Bernoulli polynomial. 

\bibliographystyle{apsrev4-2}
 \bibliography{biblio.bib}

\end{document}